\documentclass[10pt,journal,compsoc]{IEEEtran}
\IEEEoverridecommandlockouts
\usepackage[pdftex]{graphicx}
\usepackage{epstopdf}
\usepackage{tabu}
\usepackage{amsmath,amssymb,amsfonts,amsthm}
\usepackage{graphicx}
\usepackage{textcomp}
\usepackage{xcolor}
\usepackage{cite,url}
\usepackage{booktabs} 
\usepackage{multirow} 
\usepackage{amsmath}
\usepackage{bm}       
\usepackage{algorithm}
\usepackage{algorithmicx}
\usepackage{algpseudocode}
\usepackage{balance}
\usepackage{setspace}

\usepackage{caption}
\newtheorem{observation}{\textbf{Observation}}
\newtheorem{theorem}{Theorem}
\newtheorem{definition}{Definition}

\renewcommand\arraystretch{1.2}
\newcolumntype{C}[1]{>{\centering\arraybackslash}p{#1}}

\begin{document}

\title{
On-edge Multi-task Transfer Learning: Model and Practice with Data-driven Task Allocation
}
  
\author{Zimu Zheng\IEEEauthorrefmark{2}, Qiong Chen\IEEEauthorrefmark{2}, Chuang Hu, Dan Wang, \IEEEmembership{Senior~Member, IEEE,} and~Fangming~Liu\IEEEauthorrefmark{1},~\IEEEmembership{Senior~Member,~IEEE}

\IEEEcompsocitemizethanks{
\IEEEcompsocthanksitem Q. Chen and F. Liu are with the National Engineering Research Center for Big Data Technology and System, Key Laboratory of Services Computing Technology and System, Ministry of Education, School of Computer Science and Technology, Huazhong University of Science and Technology, Wuhan, China. E-mail: \{qiongchen, fmliu\}@hust.edu.cn. 
\IEEEcompsocthanksitem Z. Zheng is with the Edge Cloud Innovation Lab, Technical Innovation Department, Cloud BU, Huawei Technologies Co.Ltd, Shenzhen, China. E-mail: zimu.zheng@huawei.com.

\IEEEcompsocthanksitem C. Hu and D. Wang are with the Department of Computing, Hong Kong Polytechnic University, Kowloon, Hong Kong. E-mail: \{cschu, csdwang\}@comp.polyu.edu.hk.}

\thanks{\IEEEauthorrefmark{2} Authors contributed equally. \IEEEauthorrefmark{1}The corresponding author is Fangming Liu. }
}



\markboth{IEEE TRANSACTIONS ON Parallel and Distributed Systems,~VOL. 31, NO. 6, JUNE 2020}%
{Shell \MakeLowercase{\textit{et al.}}: Data-driven Task Allocation on the Edge}

\IEEEtitleabstractindextext{
\begin{abstract}
On edge devices, data scarcity occurs as a common problem where transfer learning serves as a widely-suggested remedy. Nevertheless, transfer learning imposes heavy computation burden to the resource-constrained edge devices. Existing task allocation works usually assume all submitted tasks are equally important, leading to inefficient resource allocation at a task level when directly applied in Multi-task Transfer Learning (MTL). To address these issues, we first reveal that it is crucial to measure the impact of tasks on overall decision performance improvement and quantify \emph{task importance}. We then show that task allocation with task importance for MTL (TATIM) is a variant of NP-complete Knapsack problem, where the complicated computation to solve this problem needs to be conducted repeatedly under varying contexts. To solve TATIM with high computational efficiency, we propose a Data-driven Cooperative Task Allocation (DCTA) approach. Finally, we evaluate the performance of DCTA by not only a trace-driven simulation, but also a new comprehensive real-world AIOps case study which bridges model and practice via a new architecture and main components design within AIOps system. Extensive experiments show that our DCTA reduces 3.24 times of processing time, and saves 48.4\% energy consumption compared with the state-of-the-art when solving TATIM.
\end{abstract}

\begin{IEEEkeywords}
Edge computing, transfer learning, data-driven task allocation, real-world application.
\end{IEEEkeywords}}

\maketitle

\section{Introduction}

Nowadays, computationally intensive machine-learning applications such as image recognition are becoming popular on resource-constrained edge devices (e.g., intelligent camera). While enjoying the merits of these applications, users are also frustrated when striking the balance between execution time and resource consumption on the edge. To address this problem, many task partitioning approaches have been proposed. Generally, an edge application is partitioned into a set of tasks which can be executed on the edge devices. For example, the video analytics application usually consists of several tasks (e.g., face detection and action classification), and allocates these tasks to multiple edge nodes to execute. Application partition and task allocation reduce the burden of a single edge device and jointly improve the performance of the application.

However, in major edge computing systems, we often face challenges in learning under data scarcity, due to either prohibitive cost (e.g., privacy concern, storage limitations, and networking costs), or inherent difficulty in obtaining required proper training samples with respect to the system complexity and uncertainty on the edge. 
Recently, \emph{transfer learning} shows its effectiveness to tackle the data scarcity issue~\cite{hutchinson2017overcoming} and serves as a widely-suggested remedy for different industrial applications with insufficient samples, e.g., image recognition~\cite{yuan13}, speech analysis~\cite{wu15}, disease diagnosis~\cite{emrani17}, medical informatics~\cite{zhou11} and industrial operations (e.g., AIOps)~\cite{ide2017multi}.

In this paper, we focus on the \emph{Multi-task Transfer Learning (MTL)} on the edge, where a machine-learning-based application can be divided into multiple machine-learning tasks, and each task can obtain the knowledge of some other tasks to improve its performance. It is well known that the machine-learning-based application is highly computation-intensive, while the computation resource of edge device is limited. Many efforts have been devoted to designing task allocation mechanisms to achieve various objectives, e.g., optimizing the makespan~\cite{biswas2017multi}, throughput~\cite{hong2007adaptive} or reliability~\cite{jiang2015reliable} of the application. However, these frameworks focus on general parallel tasks in the centralized datacenter, where the computation capacity is assumed to be infinite in terms of constantly leasing of virtual machines. 

In edge computing systems, it is sometimes hard to obtain a satisfactory result within time and resource limitations if we directly utilize existing frameworks for the cloud. Admittedly, existing task allocation studies have considered that different tasks may require different resources in edge computing systems in order to jointly improve the performance of the application~\cite{sundar2018offloading,cao2015energy,mao2016power}. They are usually designed for general machine learning and typically assume that all tasks contribute identically to overall performance improvement of the application. However, in MTL, tasks belonging to the same machine-learning-based application usually have different potential for improving the application's overall performance. Directly applying these techniques leads to inefficient resource utilization at a task level under MTL in edge computing systems.

To solve the above inefficiency issue for multiple-task allocation in edge computing systems, the key is that more important tasks, which have the higher potential for improving the application's overall decision performance, should be allocated to more powerful edge devices for priority execution under time limits. Recently, Geng et al. also considered the priority of tasks by leveraging the dependency of tasks in task allocation~\cite{geng2018energy}. In that study, the task dependency is predefined and remains fixed over time, e.g., installing Hadoop before Spark. However, due to the complex nature of machine-learning tasks, variables such as environmental conditions and model configurations are likely to change over time. The dependency of machine-learning tasks is dynamic and usually not available before learning. Directly applying the current allocation mechanism can easily result in significant overall application performance degradation for MTL on the edge.

Instead of assuming that all tasks contribute identically to the application's overall decision performance improvement and conducting the time-dynamic task allocation on the edge, our idea is to leverage machine learning techniques to capture the correlated and collective potential improvement of multiple tasks. Accordingly, we propose a Data-driven Cooperative Task Allocation (DCTA) mechanism to maximize the application's overall decision performance among multiple tasks on the edge. We also conduct a new comprehensive case study under the real-world industrial operation (e.g., AIOps) scenario, where MTL is necessary due to the data scarcity on edge devices.

\textbf{Challenges and solutions.} In designing DCTA, we have to overcome three following major technical challenges.

First, the metric of tasks impact on overall decision performance improvement remains unknown in current studies. To tackle the challenge, we propose a metric of \emph{task importance}, which is to measure the overall performance degradation when the measured task is not conducted in MTL. We also observe the long-tail property of task importance, i.e., only a few tasks are important, which serves as a key metric to guide task allocation and facilitate resource saving from less important tasks. We formally define the TATIM problem of task allocation with task importance for MTL on the edge.

Second, the TATIM problem is challenging due to not only its computation complexity (i.e., NP-complete) but also the varying contexts (i.e., dynamic task importance) on the edge. We first prove that TATIM is a variant of Knapsack problem and thus NP-complete. We then show that the task importance is difficult to capture, due to varying environmental conditions and configurations. Therefore, the complicated computation to solve this problem needs to be conducted repeatedly under varying contexts on the edge. To enhance the computational efficiency, we propose a data-driven task allocation mechanism based on reinforcement learning.
   
Third, applying the machine learning technique to solve the TATIM problem introduces a trade-off between accuracy and cost. On one hand, an accurate data-driven model requires a huge amount of expensive local data on real-world operations. On the other hand, merely using general data from simulation helps to reduce the amount of local data needed but leads to low accuracy. To tackle the challenge, we propose a cooperative learning mechanism to reduce the amount of data needed to generate a reliable data-driven model, by leveraging both general simulated data and local real-world data. 



We implement DCTA as a task allocation approach within a data-driven building management system. We also evaluate various distinct task allocation approaches by not only a trace-driven simulation, but also a new comprehensive real-world AIOps case study which bridges model and practice via a new architecture and main components design within AIOps system. Extensive experiments show that our DCTA reduces 3.24 times of processing time, and saves 48.4\% energy consumption when solving TATIM compared to the state-of-the-art.

The rest of the paper is organized as follows. In Sec. \ref{Sec:Motivation}, differing from our preliminary work \cite{chen2019}, we reorganize all the notations and observations on task importance for better understanding. In Sec. \ref{Sec:Model}, we introduce the data-driven approach for task allocation, by leveraging both cluster reinforcement learning and support vector machine. In Sec. \ref{Sec:Evaluation}, we conduct trace-driven simulations to evaluate the performance of the proposed DCTA mechanism. In Sec. \ref{Sec:CaseStudy}, we add a new comprehensive case study on AIOps for our DCTA mechanism to bridge model and practice. Specifically, we first elaborate the background of AIOps system for better understanding, and exhaustively analyze the motivation of applying DCTA mechanism to the AIOps system. We then further analyze how to apply the DCTA mechanism by proposing a new architecture and main components design within AIOps system. Extensive experiments are complemented to demonstrate the superiority of AIOps system integrating our DCTA mechanism. Sec. \ref{Sec:Related-work} discusses related work and Sec. \ref{sec:Discussion} analyzes some future work and possible improvements. At last, we conclude this paper in Sec. \ref{Sec:Conclusion}.
\section{Background and Problem Definition of Task Allocation with Task Importance} \label{Sec:Motivation}

In this section, we first introduce the background of Multi-task Transfer Learning (MTL). We then give a formal definition of task importance. We also observe the long-tail property of task importance and the potential of leveraging task importance for task allocation in MTL. With these notations, we formally define the problem of task allocation with task importance for MTL. 

\subsection{Background of Multi-task Transfer Learning (MTL)} 

\begin{figure}[t]
\hspace{-0.2cm}
\includegraphics[angle=0, width=0.5\textwidth]{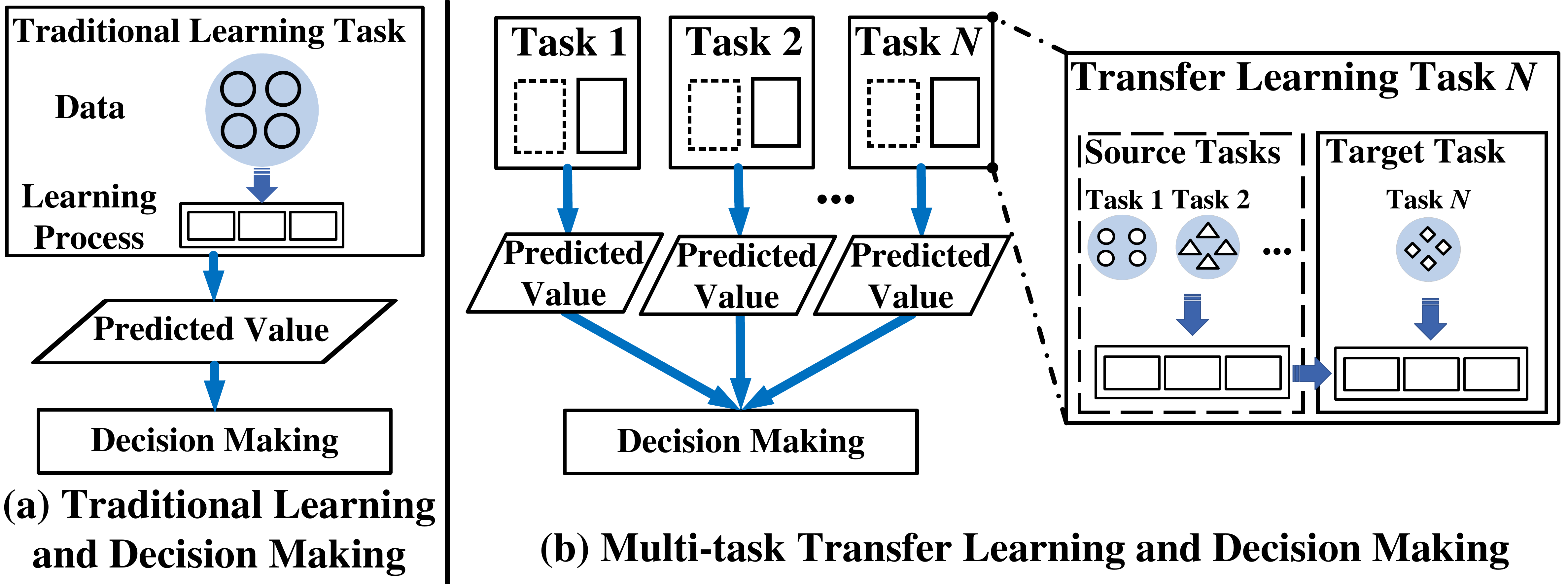}
\vspace{-0.6cm}
\caption{Decision making with (a) traditional learning and (b) transfer learning.}
\label{fig:transfer1}
\vspace{-0.5cm}
\end{figure}

In this paper, we study the issue of \emph{Multi-task Transfer Learning (MTL)} on the edge, where varying tasks together can facilitate better decision performance. It basically reuses parameters or training samples of source tasks to support target tasks, e.g., which are lack of training data. The term \emph{task} is defined as a set of data, label and its corresponding learning model for a predefined context. For example, for a self-driving car on the road, the detection of each type of object, e.g., neighboring-car, traffic-sign, or pedestrian detection, can be modeled separately as a task. Another example is to take the coefficient of performance (COP) prediction of a chiller for one particular operation as a task \cite{zheng2018data}. The process is shown in Fig.~\ref{fig:transfer1}.


The benefits of multiple tasks come in mainly two ways. First, similar tasks can transfer their knowledge between each other during the training process, which reduces the negative effect of data scarcity, especially on the edge. Second, in the real-world scenario, it is common to make the final decision by aggregating the output of multiple tasks. Maintaining the high performance of all these tasks contribute to the final aggregated decision performance. Again in the example of a self-driving car, the final driving operation of the car is conducted based on the result of multiple data-driven tasks, e.g., the neighboring-car, traffic-sign, and pedestrian detection. 

\textbf{The Computation Challenge.} However, the current MTL systems are way too computationally complicated for edge devices. The reason is twofold: 
1) Each task needs to be learned individually from scratch, where siloing tasks make training a new task or a comprehensive perception system a Sisyphean challenge; 
2) To avoid data-driven task model being out-of-date and leverage the latest accumulated data as effectively as possible, MTL practitioners retrain their models repeatedly to get the final model with the best quality, including to explore feature representation \cite{yang16,lin16,gong12}, adjust structures of task relationship~\cite{zhang17,lin16Interactive,oyen12} and tune hyper-parameters~\cite{isele16}. For better understanding, a formal formulation of MTL tasks on the edge is available in the following Section~\ref{Sec:transfer}.

\subsection{Notations of Task Importance}\label{Sec:notation}

Confronted with the computational challenge of MTL, we aim to allocate tasks for more efficient MTL on the edge. 
When allocating tasks, current studies usually assume that all machine-learning tasks are equally important so that resources should be allocated to ensure the accuracy of all these tasks.

However, tasks are not always related to the current context, and thus not equally important. At a specific period of time, e.g., within one hour, the number of highly important tasks are likely to be of a minor, compared with the number of all possible tasks. For example, for a self-driving car on the high way, neighboring car detection can be much more related and important compared with most tasks like pedestrian detection which are more important in a downtown area. \footnote{As a further demonstration, a real-world experiment and the corresponding observation are also available at the end of the subsection.}

{In this part, before studying \emph{task importance}, we first formally define it and its related notations. The key notations in this paper are also listed in Table \ref{table:notations} for ease of reference. A further experiment on task importance is available after the definition.}  



\begin{table}[!t] 
\small
\renewcommand{\arraystretch}{1.2}
\centering \caption{{List of Key Notations.}}
\label{table:notations}
\begin{tabular}{|c|c|}
\hline
Notation & Description \\
\hline \hline
$\bm{J}$ & Set of tasks where $\bm{J} = \{j\}$ \\
\hline
$\bm{P}$ & Set of edge devices where $\bm{P} = \{p\}$ \\
\hline
$\mathcal{I}_j$ &  The importance of task $j$ \\
\hline
$\mathcal{H}(\cdot)$ & The merit function indicates the ability to \\
& provide credible decision performance \\
\hline
$\mathcal{D}(\cdot)$ & The decision-making function indicates  \\
& the best operation \\
\hline
$D$ & The ideal performance \\
\hline
$u_{j,p}$ & Whether the task $j$ is assigned to  \\
& processor $p$ (=1) or not (=0) \\
\hline
$t_j$ &  The execution time of task $j$  \\
\hline
$v_j$ & The resource (e.g., battery) consumption of task $j$ \\
\hline
$T$ & The maximum time limits to conduct the decision\\
\hline
$V_p$ & The maximum resource capacity of processor $p$ \\
\hline
$\theta_j$ & The model parameters of task $j$\\
\hline
$L_j(\cdot)$ & The learning loss of task $j$\\
\hline
$\textbf{u}$ & The task-allocation matrix where $\textbf{u}$ = $[u_{j,p}]$ \\
\hline
\end{tabular}
\vspace{-8pt}
\end{table}

  
\begin{definition} (Task Importance)
Given a task set $\bm{J} = \{j\}$ which consists of a series of tasks, the importance of task $j$ is
\begin{equation} \label{eq:importance}
\mathcal{I}_j = \mathcal{H}(\bm{J};\bm\theta) - \mathcal{H}(\bm{J} \backslash \{j\};\bm\theta \backslash \{\theta_j\}), 
\end{equation}
\noindent where a learning task is denoted by $j \in \mathbb{N}^+$; $\theta_j$ denotes the model parameters of task $j$ and $\bm{\theta} = \{\theta_j\}$ denotes its vector; merit function $\mathcal{H}(\cdot)$ outputs the final potential performance improvement; $\bm{J}$ denotes the entire task set.
\end{definition}

Thus, given model parameters $\bm\theta$, the task importance $\mathcal{I}_j$ can be updated using the merit function $\mathcal{H}(\cdot)$. {Such a function indicates the ability to provide credible decision performance (e.g., energy saving) and outputs a value called \emph{overall merit}, which is formally defined as below.} 

\begin{definition} {(Overall Merit) Given the task set $\bm{J}$ and the ideal performance of final decisions $D$, the overall merit is defined as the similarity with the ideal performance, i.e., 
\begin{eqnarray}\label{Eq:merit}
\emph{OM} = \mathcal{H}(\bm{J};\bm\theta) = 1- \frac{|D - \mathcal{D}(\bm{J};\bm\theta)|}{D},
\end{eqnarray}
\noindent where $\mathcal{D}(\cdot)$ denotes a decision-making function given model parameters, and $D$ denotes the ideal performance which can usually be collected after final optimization, i.e., collected manually or automatically by leveraging historical samples.} 
\end{definition}

{In general, the historical data records the descriptor of contexts/ scenarios, requirement, historical operations, and its results. Such information helps us to define $D$. For example, in the case of a self-driving car, in order to ensure the car arrive at the destination safely, it will conduct a series of decision actions where the least time-consuming situation can be regarded as the ideal performance.}



Such a decision-making function $\mathcal{D}(\cdot)$ is intrinsically solving an optimization problem finding the best action according to parameters, which can be set once given the scenario. For example, in the case of a self-driving car, a possible decision-making function is to find an action which minimizes the probability of accident while ensures the car should be able to arrive at the destination under time limitations. {For interested readers, a more concrete implementation of $\mathcal{D}(\cdot)$ is also available in Section~\ref{Sec:CaseStudy}.}


\begin{figure}[t]
\vspace{-0.2cm}
\begin{minipage}[t]{3.9cm}
  \includegraphics[angle=0, width=1.1\textwidth]{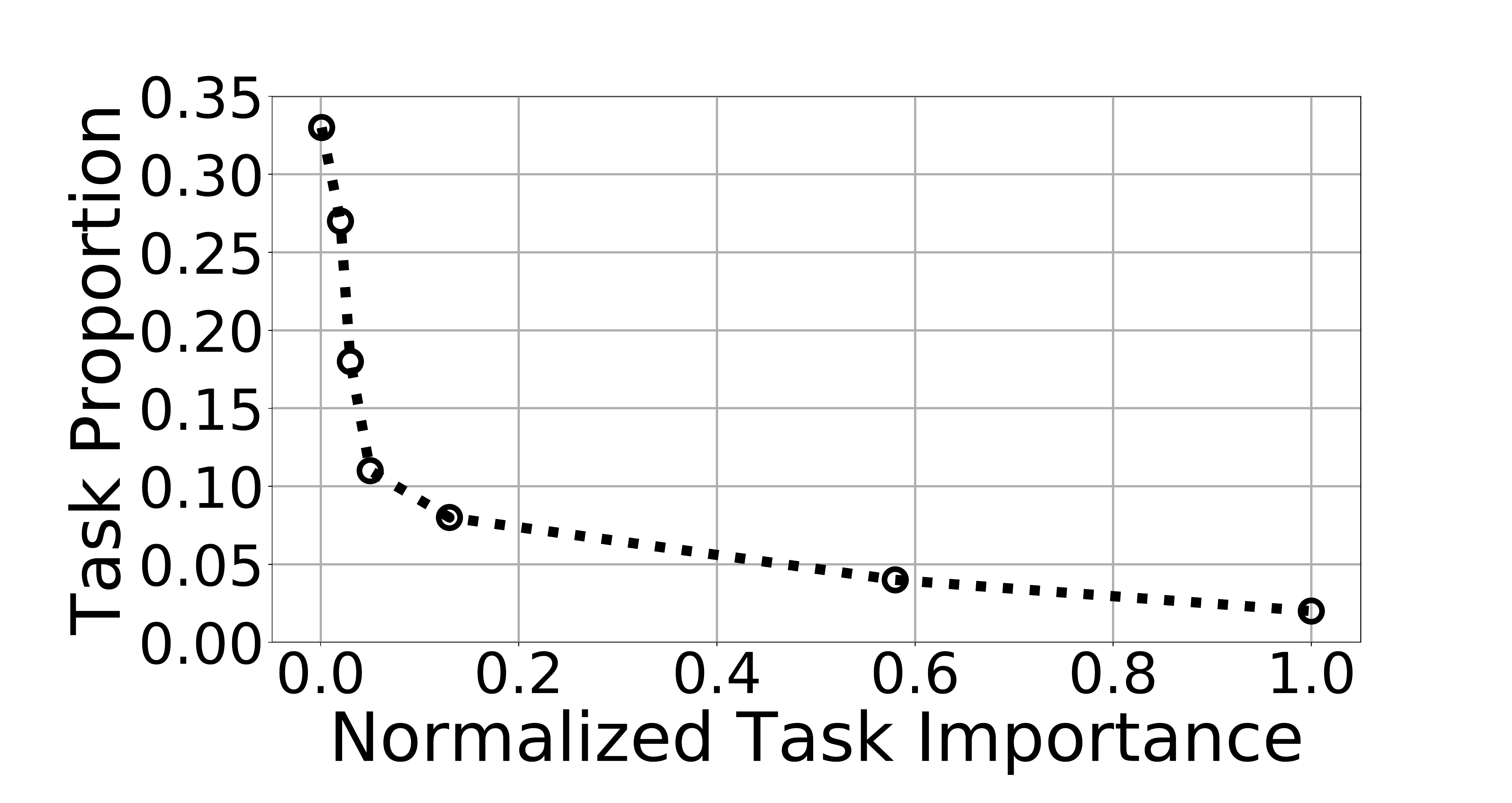}
  \vspace{-0.5cm}
  	\caption{{The non-uniform (i.e., long-tail) distribution of normalized task importance in MTL.}}
  \label{fig:importance-proportion}
\end{minipage}
\hspace{0.5cm}
\begin{minipage}[t]{3.9cm}
	\includegraphics[angle=0, width=1.1\textwidth,keepaspectratio]{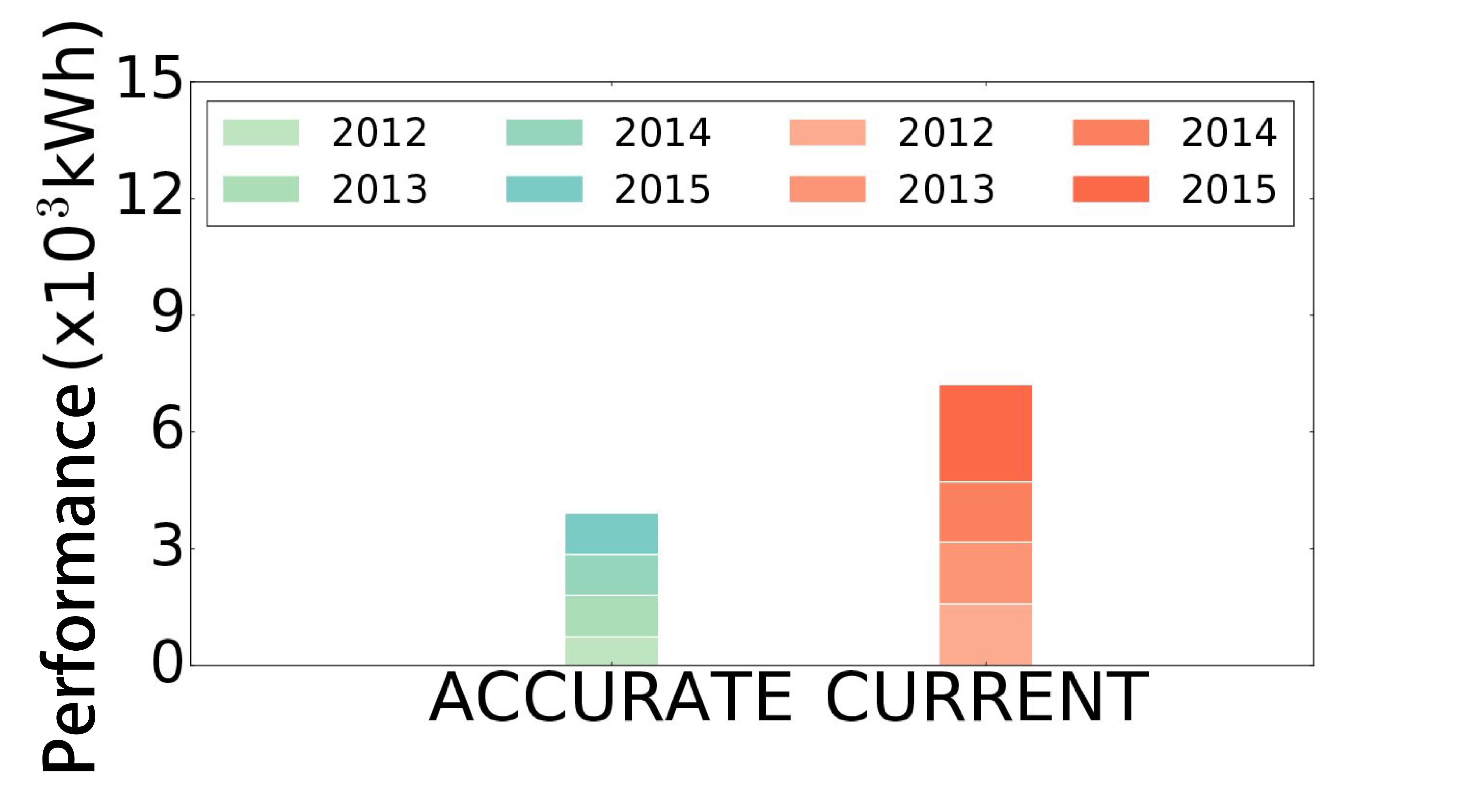}
    \vspace{-0.5cm}
    \caption{{The decision performance (i.e., energy consumption) with ACCURATE and CURRENT schemes.}} 
    \label{fig:potential-improvement}
\end{minipage}
\vspace{-0.4cm}
\end{figure}



\subsection{Observations on Task Importance}\label{Sec:observation}
{We have introduced related concepts of task importance. We next justify the motivation of using task importance by further observations.} 

We first plot the distribution of task importance in Fig.~\ref{fig:importance-proportion}, based on a real-world MTL dataset released in \cite{zheng2018data}. In there are totally 50 data-driven tasks for cooling operations running across four years in three buildings. We observe a long-tail property of task importance, i.e., merely 12.72\% of tasks have a high contribution of over 80\% to the overall merit. {We say that a task is \emph{unimportant} when its task importance is critically low compared with others, e.g., below 0.05\%.} We therefore have such an observation. 

\begin{observation}
    In MTL, unimportant tasks exist; The importance of tasks obeys a long-tail distribution.
\end{observation}

{This observation reveals the non-uniform distribution of task importance in the real-world environment which motivates us to break the common assumption of modern MTL.} Results in a recent CVPR paper also confirm such an observation~\cite{taskonomy2018}. The unimportant (e.g., redundant or noisy) tasks can be the result of 1) insufficient training samples on the edge, and 2) mismatch of context and submitted tasks in practical scenarios. {It also indicates the potential of speeding up MTL from those unimportant tasks.} 

{In the machine learning community, current MTL systems usually conduct tasks in the order of time stamps, where these time-ordered tasks are of arbitrary importance.} Thus, the current execution sequence can be regarded as random, e.g., normally distributed, in terms of task importance. When there are limitations on resource and execution time for MTL tasks, the current approach can suffer from lower overall merit.


{We conduct experiment on the MTL dataset mentioned above~\cite{zheng2018data}, where the decision objective of MTL is to control the Chiller AIOps system to minimize the energy consumption for cooling, which refers to the decision performance. Fig.~\ref{fig:potential-improvement} shows the result by conducting MTL tasks in the order of task importance (called \emph{ACCURATE} scheme), compared with the order of time with random task importance (called \emph{CURRENT} scheme) under execution time limitations. Such an ACCURATE scheme can be obtained by computing task importance using historical data (Section~\ref{Sec:notation}), and can be regarded as ground truth. Base on the obtained accurate task importance, we can find the best task allocation strategy. For interested readers, a detailed optimization process is also available in Section~\ref{SubSec:5.2}.} 

{Stacked bars on the left indicate the performance with the ACCURATE scheme, whereas the right show the CURRENT scheme using random task allocation. We see that the ACCURATE scheme considering task importance could have resulted in an average of over 45.68\% potential improvement in terms of the overall merit. These results demonstrate that there is significant room to improve the overall merit when using a more accurate and robust scheme of task allocation. We summarize the observation as below.}



\begin{observation}
    {Overall merit with MTL can be improved by task allocation according to task importance.}
\end{observation}


{However, the task importance may not be always directly available for run-time usage. The above experiment is based on historical data so that we are able to compute the task importance after a task is executed. For run-time usage, we often need to know the task importance in advance, i.e., before a task is conducted. A natural question is whether the task importance is easy to predict, e.g., a fixed or stable value.} Based on the above MTL dataset, we also conduct two experiments as more-detailed distribution studies showing how the importance fluctuates over operations under different industrial demands and conditions. 

\begin{figure}[t]
\vspace{-0.2cm}
\begin{minipage}[t]{3.9cm}
  \includegraphics[angle=0, width=1.1\textwidth]{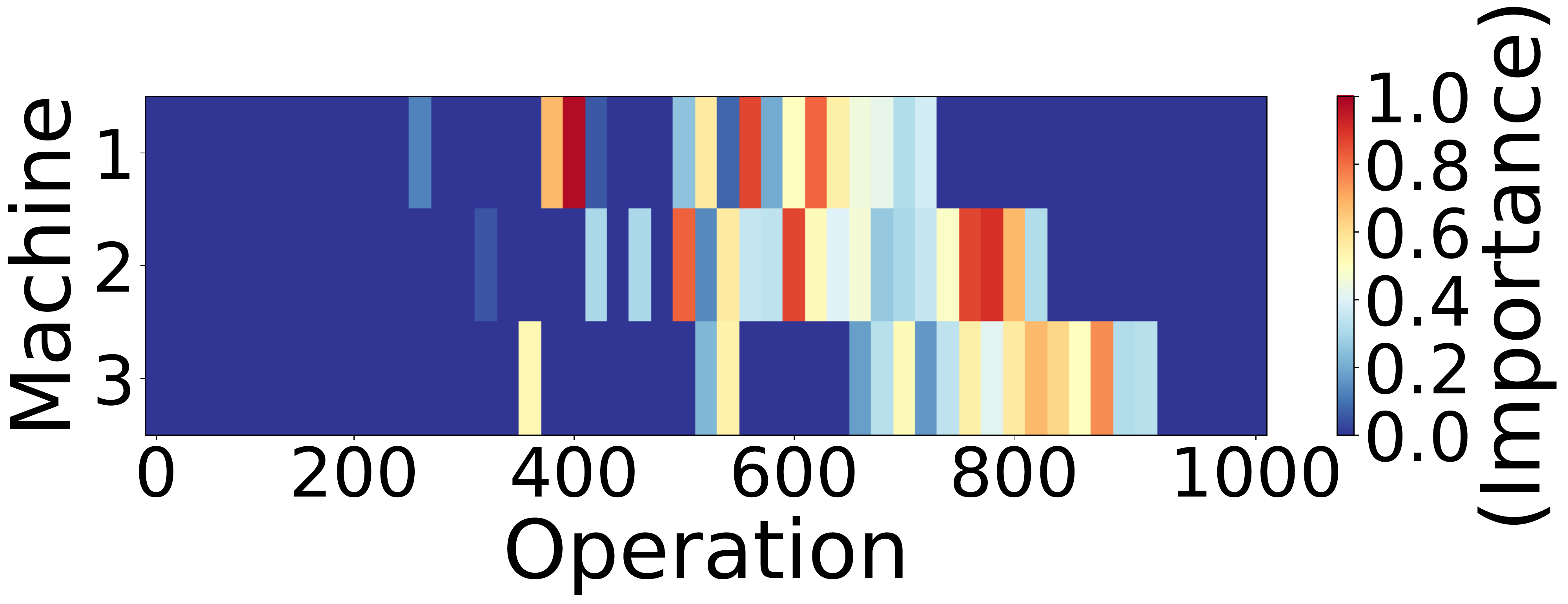}
  \vspace{-0.5cm}
  	\caption{Average task importance for different types of machines and operations.}
  \label{fig:avgtask2}
\end{minipage}
\hspace{0.5cm}
\begin{minipage}[t]{3.9cm}
	\includegraphics[angle=0, width=1.1\textwidth,keepaspectratio]{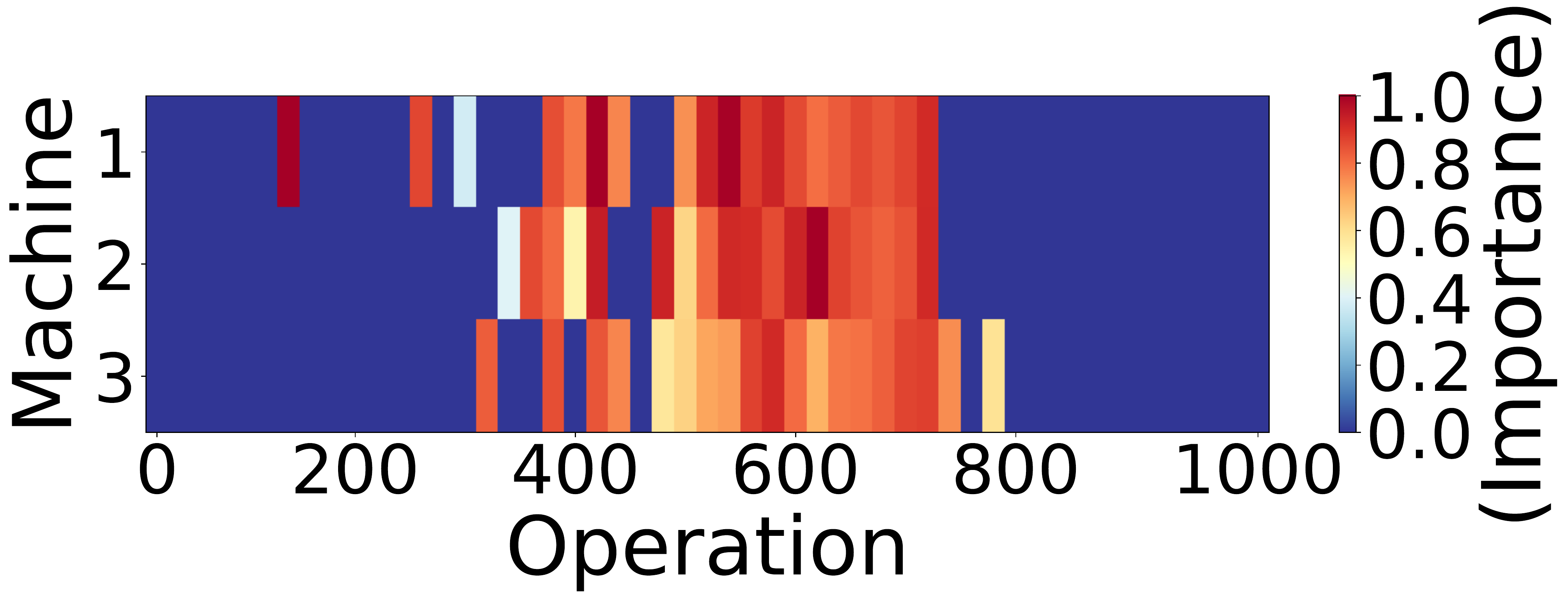}
    \vspace{-0.5cm}
    \caption{Task importance variation for the same types of machines and operations.} 
    \label{fig:vartask2}
\end{minipage}
\vspace{-0.5cm}
\end{figure}

{We first plot the average task importance as a function of different operations in terms of different types of machines in Fig. \ref{fig:avgtask2}. We pick the first regular machine for example. It can be seen that these machines often operate at a small portion of operations, and the importance fluctuates somewhat randomly. At the same time, for the same types of machines, we plot in Fig. \ref{fig:vartask2}, the variation in their task importance under different operations, and note that there is a large fluctuation even for a given operation. This is because the task importance in practice is highly dependent on a variety of factors like environmental conditions and configurations. Such factors are referred as the term \emph{context} in this paper.} We therefore have such an observation.


\begin{observation}
    {Task importance fluctuates markedly over varying contexts with MTL in terms of average and variance.}
\end{observation}

This observation reveals that the time-dynamic task importance changes in varying contexts~\cite{hu2018synthesize,sax2018mid}, e.g., with different external factors (like environmental conditions and dynamic industrial demands) and internal factors (like machine configurations and response). 
{For example, in the case of self-driving car, the \emph{context} contains the following specific factors such as visual observations, physical information, weather, traffic conditions and etc. These factors are exceedingly difficult to capture within an analytical model. Facing such a high variance of task importance situation, natural thinking of modeling task importance using synthetic models easily suffers from low accuracy.}


\subsection{Problem of Task Allocation with Task Importance for MTL} 
\label{Sec:transfer}
 

{Based on the above notations and observations, the intuition behind this paper is that we should allocate more important tasks to more powerful edge devices (e.g., edge server) to optimize the final decision. Here we say an edge device is more powerful which refers to it's processing speed or frequency is faster.} We aim to leverage task importance to facilitate task allocation for MTL tasks on the edge, with an emphasis of time limits.

We start by formally defining the conception of \emph{task allocation} and \emph{MTL tasks on the edge}, {where the former consists of the task placement and resource allocation. Considering machine-learning tasks are usually highly computation intensive, resource-constrained edge devices can barely handle multiple tasks in parallel. Therefore, we assume that, at a certain time, a \emph{task} occupies the whole CPU computing resource under execution.}

\begin{definition} (Task Allocation)
Given an edge device set $\bm{P} = \{p\}$ which consists of a series of edge devices, the task allocation over $\bm{P}$ is a binary variable $u_{j,p}$, i.e.,
\begin{eqnarray*}
u_{j,p} = 
\begin{cases} 
1, & \mbox{if task $j$ is assigned to edge device $p$} 
\\ 0, & \mbox{otherwise,}
\end{cases} 
\end{eqnarray*} 
where an edge device is denoted by $p \in \mathbb{N}^+$.
\end{definition}

Since each task is indivisible and must be assigned to exactly one edge device, we have the following constraint:
\begin{equation}\label{Eq:2}
\sum \limits_{p \in \bm{P}} u_{j,p}= 1,\ \forall j \in \bm{J}. 
\end{equation}


{Considering edge devices are usually resource-constrained and discrete, we classify resources into two categories, i.e., execution-related and basic requirements. The former refers to CPU computing resources, whereas the latter refers to battery or storage resources. Therefore, the CPU execution time and basically resource requirements of all tasks assigned to edge device $p$ should satisfy the following constraints:}

\begin{equation}\label{Eq:3}
\sum \limits_{j \in \bm{J}} t_j \cdot u_{j,p} \leq T, \ \ \forall p \in \bm{P},
\end{equation}
\begin{equation}\label{Eq:4}
\sum \limits_{j \in \bm{J}} v_j \cdot u_{j,p} \leq V_p, \ \ \forall p \in \bm{P},
\end{equation}
where $t_j$ denotes the execution time of task $j$; $T$ denotes the time limitations; $v_j$ denotes the resource required for task $j$; $V_p$ denotes the resource capacity of edge device $p$.


The objective of traditional MTL is to minimize the collective loss of all tasks. We study the modeling and define the MTL tasks specific to the edge computing scenario for better understanding.

\begin{definition} (MTL Tasks on the Edge) \label{def:mtl}
    Given task importance $\mathcal{I}_j$, the execution time and resource limitations of Eq. (\ref{Eq:2})~-~(\ref{Eq:4}), an on-edge MTL tasks aims to obtain $\theta$ by
    \begin{equation*}
    \begin{aligned}
    \bm{\theta} = \arg \min \sum_{j \in \bm{J}} \sum_{p \in \bm{P}} \mathcal{I}_j \cdot L_j(\theta_j) \cdot u_{j,p}, \ s.t. \  Eq. (\ref{Eq:2}) - (\ref{Eq:4}), 
    \end{aligned}
    \end{equation*}
    where $L_j(\theta_j)$ denotes the learning loss of task $j$, e.g., prediction error and regularization terms.
\end{definition}

Based on the above definitions, we formally define the problem of task allocation with task importance for MTL on the edge (TATIM Problem) as below.

\begin{definition} (TATIM Problem) Given the execution time and resource limitations, a TATIM problem is to obtain \textbf{u} by 
    \begin{equation*}\label{Eq:allocation}
    \begin{aligned}
    \max   \limits_{\textbf{u}} 
      \sum_{j \in \bm{J}} \sum_{p \in \bm{P}} \ \mathcal{I}_j \cdot u_{j,p}, 
    \mbox{ s.t.}
    \ Eq. \ (\ref{Eq:2}) - (\ref{Eq:4}),
    \end{aligned}
    \end{equation*}
    where $\textbf{u}$ = $[u_{j,p}]$ denotes the task-allocation matrix; $\mathcal{I}_j$ can be computed given $\bm\theta$ from Definition~\ref{def:mtl} and $\bm J$ using Equation~\ref{eq:importance}$-$~\ref{Eq:merit}.
\end{definition}

We found that the TATIM problem under the execution time and resource limitations is in fact a 0-1 Knapsack problem, which is in general NP-complete. 

\begin{theorem}\label{theorem:1}
Task allocation problem with task importance is a 0-1 multiply-constrained multiple Knapsack problem.
\end{theorem}

{For interested readers, the proof of Theorem~\ref{theorem:1} can be found in our conference paper \cite{chen2019}.}

\section{Data-driven Approach for Task Allocation}\label{Sec:Model}

{As shown in the previous section, when we introducing the time-varying task importance $\mathcal{I}$, task allocation becomes a TATIM problem which is challenging as an NP-complete problem twofold.}

{First, the complexity introduced by task importance is the reason why we adopt a reinforcement learning (RL) model. We leverage data-driven methods in order to reduce the time needed to solve the origin NP-complete Knapsack problem. Specifically, in the data-driven RL method, we integrate task importance into the environment modeling of RL.}

{Second, because the task importance is time-varying, an RL model cannot simply be applied. In the first part, we propose a clustered reinforcement learning (CRL) model that makes decisions based on how observations of the environment relate to those previously seen. In the second part, because the CRL model can confront with quite a few unseen environments, we further propose a Support Vector Machine (SVM) model to predict the task importance and dynamically adjust CRL model decisions based on real-time data.}  

{In a brief summary, the reason for using data-driven technique for TATIM with task importance is because it shows its effectiveness for complicated problems in time-varying environments, including Intelligent logistics \cite{li2018development}, Autonomous Mobility-on-Demand system \cite{iglesias2018data}, and Human-level game control \cite{mnih2015human}. Basically, data-driven techniques are particularly helpful for solving complicated problems repeatedly with varying parameters, because they not only help to model and reduce the environmental randomness in multi-task scenarios but also help to significantly enhance the computational efficiency due to the fast inference phase when the solution is needed.\footnote{Though the training phase may be long, it merely needs to be conducted once in advance.}}

Formally, given a task set $\bm{J}$ and the corresponding historical feature space $\mathcal{X}$, we are to develop a data-driven task allocation scheme with a loss function $\mathcal{L(\cdot)}$ which maximize the overall decision performance of the task allocation, i.e.,
\begin{equation*}
\begin{aligned}
\textbf{u} \leftarrow \mathcal{F}(\bm{J},\mathcal{X}).
\end{aligned}
\end{equation*} 
 

\subsection{The Clustered Reinforcement Learning (CRL) Model}

Next, we consider the proper approach to solving the TATIM problem. 
First, in the previous section, we have proved that the TATIM problem is in fact a Knapsack problem and therefore NP-complete. RL is widely suggested to efficiently solve such problems \cite{iglesias2018data,mnih2015human}. 
Second, decisions made by industrial systems can be highly repetitive, thus generating an abundance of training data to support complicated data-driven model. 
Based on the two reasons, we applied the well-known RL to solve the TATIM problem.  

In general, the RL works like this: at each decision epoch, the agent will make a decision based on the current state of the environment. Once the decision is made, a reward would be provided to the agent and the state of the environment would be updated for making future decisions. The agent tries to maximize the cumulative rewards over time.
With RL, our TATIM problem is optimized in a Markov Decision Process (MDP), which is a five-tuple: $<\mathcal{S}, \mathcal{A}, \mathcal{P}, r, \lambda>$,  where $\mathcal{S}$ denotes the set of states; $\mathcal{A}$ denotes the set of actions; $\mathcal{P}$ denotes the transition probability distribution; $r$ denotes the reward function and $\lambda \in [0,1]$ denotes the discount factor for future rewards. Note that different optimization problems have quite different objectives, constraints, and variables. To adopt our TATIM problem, the different components of RL needs to be specially designed. The detailed design of these components in RL and MDP will be discussed next.

\textbf{Environment-dynamic Task Allocation.} However, RL should not be directly applied in our scenario, where the environment is diverse over time and existing RL approaches usually assume a fixed environment.

\textbf{1) Novel Problem of Environment-dynamic Knapsacks.} In TATIM, the task importance is critical for environment modeling and thus also important for RL. As we known, the knowledge learned by the decision of an agent is rewarded according to the environment. Once the task importance and the corresponding environment is not close to reality, the decision made by the agent will lead to poor performance. 

However, due to the varying scenarios in MTL, the environment matrix of RL usually changes over time in reality. Recall the previous example where a self-driving car on the highway and pedestrians usually do not occur, the task of pedestrians detection is less important compared to other tasks. Nevertheless, when driving around the school, pedestrians are particularly frequent which makes the task of pedestrians detection more important. Therefore, we see that the environment is clearly diverse in different scenarios, especially when the task importance is encoded in the environment of RL.\footnote{Even in the same scenario, the environment can change over time, due to the accumulating size of training data and the overwritten when the storage is insufficient. Experiments in Section~\ref{Sec:observation} also indicates the fluctuation between historical and current task importance.}


In this regard, directly leveraging the RL model can easily mismatch the environment and submitted less important tasks, which leads to poor decision performance \cite{bai2015information,hu2018synthesize}. We also conduct an experiment to demonstrate the negative impact. It shows a 46.28\% reduction of performance when the environment is not accurate using existing RL.

To this end, we realize that our TATIM problem can be regarded as a novel variant of the Knapsack problem. It is even more challenging than the Multiply-constrained Multiple Knapsack Problem proved in the previous section. This time, additionally, the item value (i.e., task importance) can be changed randomly over time, instead of being fixed in the traditional Knapsack problem.

\begin{figure}[t]
\centering
\vspace{-0.2cm}
\includegraphics[angle=0, width=0.45\textwidth]{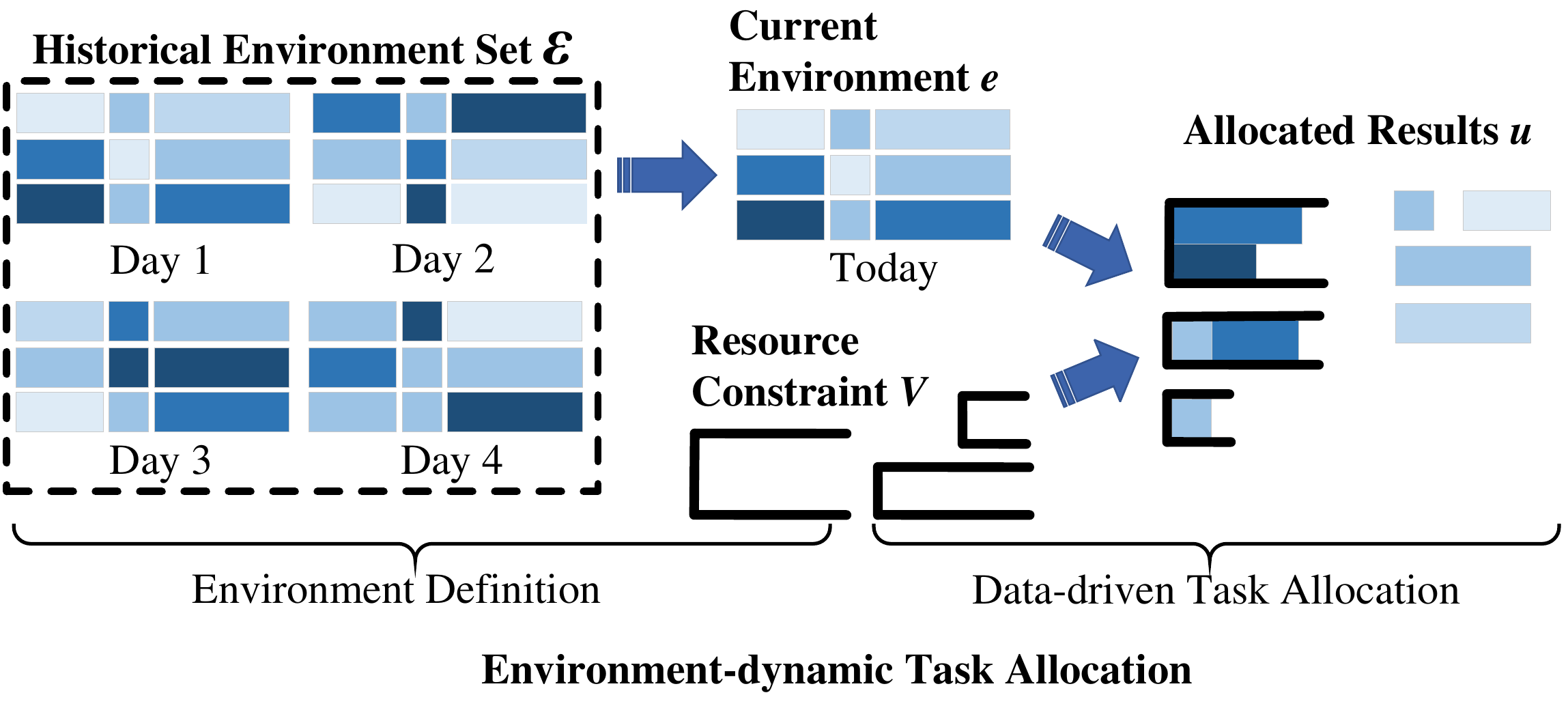}
\vspace{-0.3cm}
\caption{The illustration of environment-dynamic task allocation.} 
\label{fig:dynamic-environment}
\vspace{-0.4cm}
\end{figure}

\textbf{2) Clustered Approach for Environment Definition.} Accordingly, to solve the TATIM problem, we are to learn the current environment. Our idea is that the more similar historical days, the more similar the environment is. Such similarity can be measured by comparing the current scenarios and configuration settings, e.g., sensing data, of the predicting day and the historical days. 

The overall process is illustrated in Fig.~\ref{fig:dynamic-environment}, which consists of two parts, i.e., environment definition and data-driven task allocation. In the figure, different days represent different environments, and the darkness of each color represents the different task importance. Through the analysis of historical data, we establish an environment data set, i.e., historical environment $\mathcal{E}$. We define the historical environment $\mathcal{E}$ as the collection of environment $e$, i.e.,
\begin{eqnarray*}
\mathcal{E} = [e_{1}, e_j,\cdots,e_{N'}],\ \ \  \forall j \in [1,2,\cdots,N'],
\end{eqnarray*}
where $e_j$ denotes the corresponding environment.

Through environment definition that we can find a similar environment $e$ by clustering algorithms such as \emph{k Nearest Neighbors} (kNN), i.e.,
\begin{eqnarray*}
e = kNN(\mathcal{E},Z), 
\end{eqnarray*}
where $Z$ denotes the sensing data. We then can make data-driven task allocation based on the clustered environment under the execution time and resource constraints.

\textbf{Clustered Reinforcement Learning for Environment-dynamic Task Allocation.} Next, we propose key designs of our approach, i.e., the environment modeling, state space, action space, reward function, and optimization, which should be specified based on our TATIM problem.

\textbf{1) Environment.} A key component in the RL model is the \emph{environment}, which is everything outside the agent, and changes its state due to the action of the agent, and gives the agent corresponding rewards. For an RL predictor, the environment can be described as a matrix $e$ which is a map of the agent, e.g., Maze problem. More specifically, one dimension represents the subject types (e.g., neighboring car detection, traffic sign detection, and pedestrian detection), and the other represents the available processors (CPU processor, GPU processor, sensors). The elements of the matrix can be viewed as a data-driven task. It is formulated as follows:
\begin{eqnarray*}
e = [I_{j}\times V_p]_{N \times M},\ \ \  \forall I_{j}, V_p \in \mathbb{R},
\end{eqnarray*}
where $I_{j}$ denotes the corresponding task importance and $V_p$ denotes the corresponding processor capacity.


\textbf{2) State space.} We represent the state, which is the current task selection of the system. Specifically, the state is defined by a matrix $\mathcal{S}$ and the element of each position can be 0 or 1. Note that 1 represents the task is selected, otherwise, it is not selected, which is formulated as follows: 
\begin{eqnarray*}
\mathcal{S} = [s_{ij}]_{N \times M},\ \ \  \forall s_{ij} \in \{0,1\}.
\end{eqnarray*}
Such a fixed state representation indicates that it can be conveniently applied as an input to a neural network.

\textbf{3) Action space.} At each point in time, the scheduler may want to select any subset of the $N\times{M}$ tasks. But this requires a large action space of size $2^{N\times{M}}$ leading to unbearable computation to learn on the edge. We keep the action space small using a trick: we allow the agent to execute merely one action in each time step. The action space is given by $\{{1,2,\cdots,M}\}$, where $a$ = ${j}$ means to conduct the $j^{th}$ task for the current processor in the current time step. Hence, the action space is defined as follows:
\begin{eqnarray*}
\mathcal{A} = \{ a|a \in \{ 1,2,\cdots,M \} \}.
\end{eqnarray*}
In this way, we can greatly speed up our learning rate while keeping the action space linear in $M$.

\textbf{4) Reward Function.} We craft the reward signal to guide the agent towards desired solutions for our objective: maximize overall task importance. Specifically, we set the reward at each time step to $\sum_{j \in \bm{J}} I_j$ only if the agent reaches the terminal state (i.e., all tasks in the current system are assigned accordingly), where $\bm{J}$ is the set of tasks currently in the system. Otherwise, the reward was set to 0. Hence, 
\begin{eqnarray*}
r(t) = 
\begin{cases} 
\sum_{j \in \bm{J}} I_j, & \mbox{if the agent reaches the terminal state } 
\\ 0, & \mbox{otherwise.}
\end{cases} 
\end{eqnarray*}

It is worth noting that the agent is set to not receive any reward for intermediate decisions during a time step, which is well-suited to apply to our real-world decision objectives.

\textbf{5) Optimization.} With the above key elements, we leverage Deep Q-learning $Q(s,a; \theta, \bm{J})$ \cite{liu2017hierarchical}, where $\theta$ denotes the adjustable parameter vector of neural networks. It estimates the value of executing an action $a$ from a given state $s$. Formally, given the feature space $\mathcal{X}$ which consist of the environment $e$ and the initial state $s_0$, we have
\begin{eqnarray}
\textbf{u} \leftarrow \mathcal{F}_1(\bm{J},\mathcal{X}) = \mathcal{F}_1(\bm{J},(e,s_0)) = Q(s,a;\theta,\bm{J}).
\end{eqnarray}


Based on the above design, we propose the Clustered Reinforcement Learning (CRL) approach, as shown in Algorithm~\ref{alg:CRL}.

\begin{algorithm}
\caption{Clustered Reinforcement Learning (CRL)}\label{alg:CRL}
\begin{algorithmic}[1]
\Require
\State $\bm{\mathcal{E}} \gets$ historical environment \ \ $s_0 \gets$ initial state \ \ $Z \gets$ current scenarios and configuration settings.
\State $e \gets$ \emph{EnvironmentDefinition}($\bm{\mathcal{E}}$,$Z$) \Comment{Find similar environment.} 
\While{not yet reach the terminal state $s_N$}	 \label{alg:sc}
  \State $\mathcal{L}(s,a| \theta) \gets (r + \max \limits_{a} Q(s',a| \theta) - Q(s,a| \theta))^2$  \Comment{Update DNN parameters $\theta$.}
\EndWhile 
  \State $\bm{\theta ^*} \gets \arg \min \mathcal{L}(s,a| \theta)$  \Comment{Obtain optimal parameter $\theta$.}
 \State \Return $e, \ s_0,\ \bm{\theta ^*}$
\Ensure
\State $(e, \ s_0,\  \bm{\theta ^*})$ $\leftarrow$ initialization using the return value of the training phase.
\State $\textbf{u} \leftarrow \mathcal{F}_1((e,s_0);\bm{\theta ^*})$ \Comment{Make task allocation prediction.}
\State \Return $\textbf{u}$
\end{algorithmic}
\end{algorithm}

\subsection{The Cooperative Learning Model based on CRL} 

\begin{figure}[ht]
\centering
\vspace{-0.2cm}
\includegraphics[angle=0, width=0.42\textwidth]{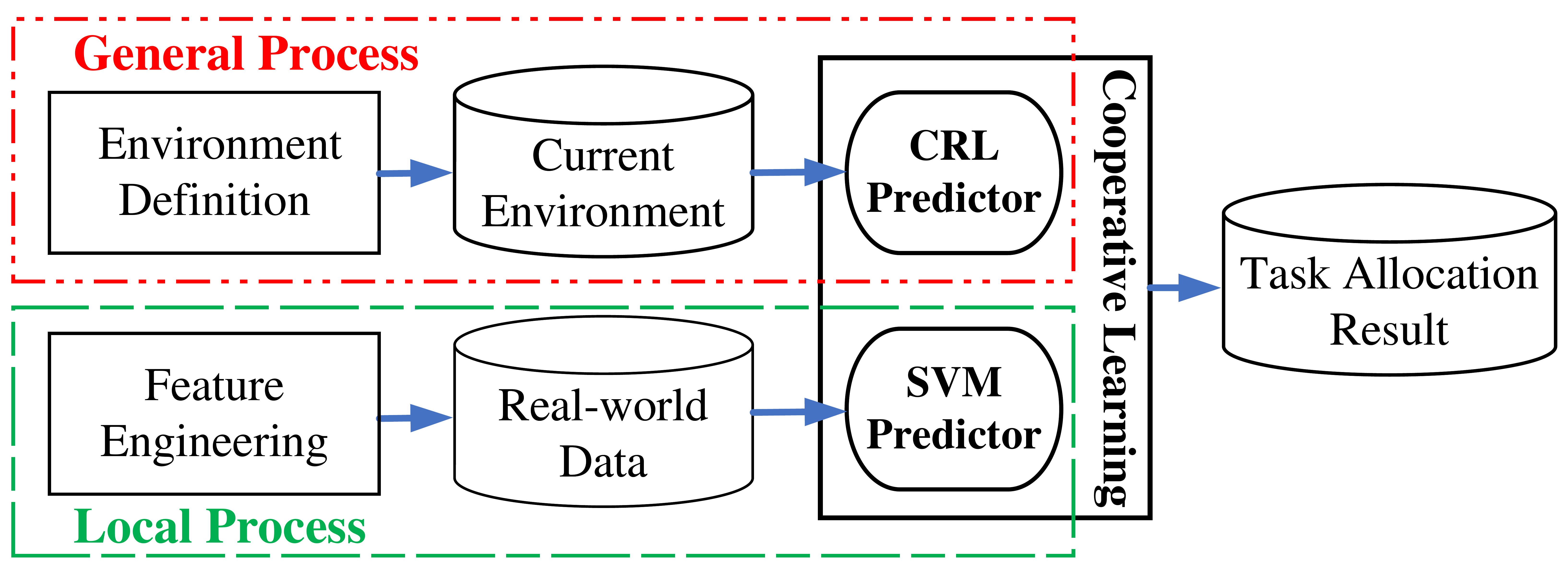}
\vspace{-0.2cm}
\caption{Framework of cooperative learning for task allocation.}
\label{fig:Clearning}
\vspace{-0.5cm}
\end{figure}

However, the CRL model should not be directly applied. In our scenario, the environment is diverse over time. Although we can find similar environments in the historical environment through simple clustering methods, there is a risk that the environment is still not closed to the real environment. That is especially true for edge devices without too much data, whereas the RL model can confront with quite a few unseen environments and it requires much environment observations to cover all possible situations.

In this regard, directly leveraging the CRL model can still mismatch the environment and submitted less important tasks, which leads to poor decision performance \cite{hu2018synthesize}. We also conduct an experiment to demonstrate the negative impact. Based on our CRL model, when the environment is not accurate, it leads to a 28.84\% reduction of performance.

\textbf{A Cooperative Learning Approach.} To tackle the challenge, our idea is to leverage runtime data to adjust the decision of the CRL model.

Accordingly, we propose a cooperative learning approach as shown in Fig. \ref{fig:Clearning}, which is especially well-suited to solve this problem. The proposed cooperative learning approach contains two components: 1) a CRL predictor with a huge environment definition data, and 2) an SVM predictor with few real-world data. Formally, let $\mathcal{C}$ and $\mathcal{R}$ be the feature spaces of the environment definition data, i.e., $\mathcal{C} = \{(e,s_0)\}$, and real-world data, respectively. Let $\mathcal{F}(\cdot)$ denotes our cooperative learning model, which can be represented more specifically as: 
\begin{eqnarray}
\mathcal{F}(\bm{J},\mathcal{X}) = \mathcal{F}(\bm{J},(\mathcal{C},\mathcal{R})) = w_1\mathcal{F}_1{(\bm{J}, \mathcal{C})}   + w_2\mathcal{F}_2{(\bm{J}, \mathcal{R})},  
\end{eqnarray}
where $\mathcal{F}_1(\cdot)$ and $\mathcal{F}_2(\cdot)$ denote the CRL predictor and SVM predictor; $w_1$ and $w_2$ denote the weight of the corresponding model results, respectively. In addition, the task-allocation matrix $\textbf{u}$ is outputted by our cooperative learning model $\mathcal{F}(\cdot)$, i.e., $\textbf{u} \leftarrow \mathcal{F}(\bm{J},(\mathcal{C},\mathcal{R}))$.

As for the SVM predictor, we compare several state-of-the-art models of SVM, AdaBoost, and Random Forest. We select SVM because of its highest accuracy. Formally, given the target tasks feature values $\mathcal{X}$, our objective is to develop an SVM predictor $\mathcal{F}_2(\cdot)$ which infers the target tasks allocation $\textbf{u}$. This can be formulated as follows:
\begin{eqnarray}
\textbf{u} \leftarrow  \mathcal{F}_2(\bm{J},\mathcal{X}) = \emph{SVM}(\mathcal{X};w,\bm{J}),
\end{eqnarray}
where $w$ denotes its parameter vector. 
\footnote{For interested readers, the design of loss function and feature engineering can be found in our conference paper \cite{chen2019}.}

\section{Performance evaluation} \label{Sec:Evaluation}

In this section, we investigate the performance of DCTA with extensive simulations over industrial operation (e.g., AIOps) scenarios  using real-world data obtained from multiple data-driven building management systems.

\subsection{Experiment Setup}

For generating MTL tasks, we use a real-world {\em building operation} dataset released in \cite{zheng2018data}, which contains four-year operation data for three high-rise commercial buildings in a metropolitan, collected by a major building service provider. The total data is more than 1 TB. Supported 50 MTL tasks include independent multi-task learning, self-adapted multi-task learning and clustered multi-task learning based on SVM, AdaBoost and Random Forest. 

\begin{figure}[t]\footnotesize
\centering
\includegraphics[angle=0, width=0.3\textwidth]{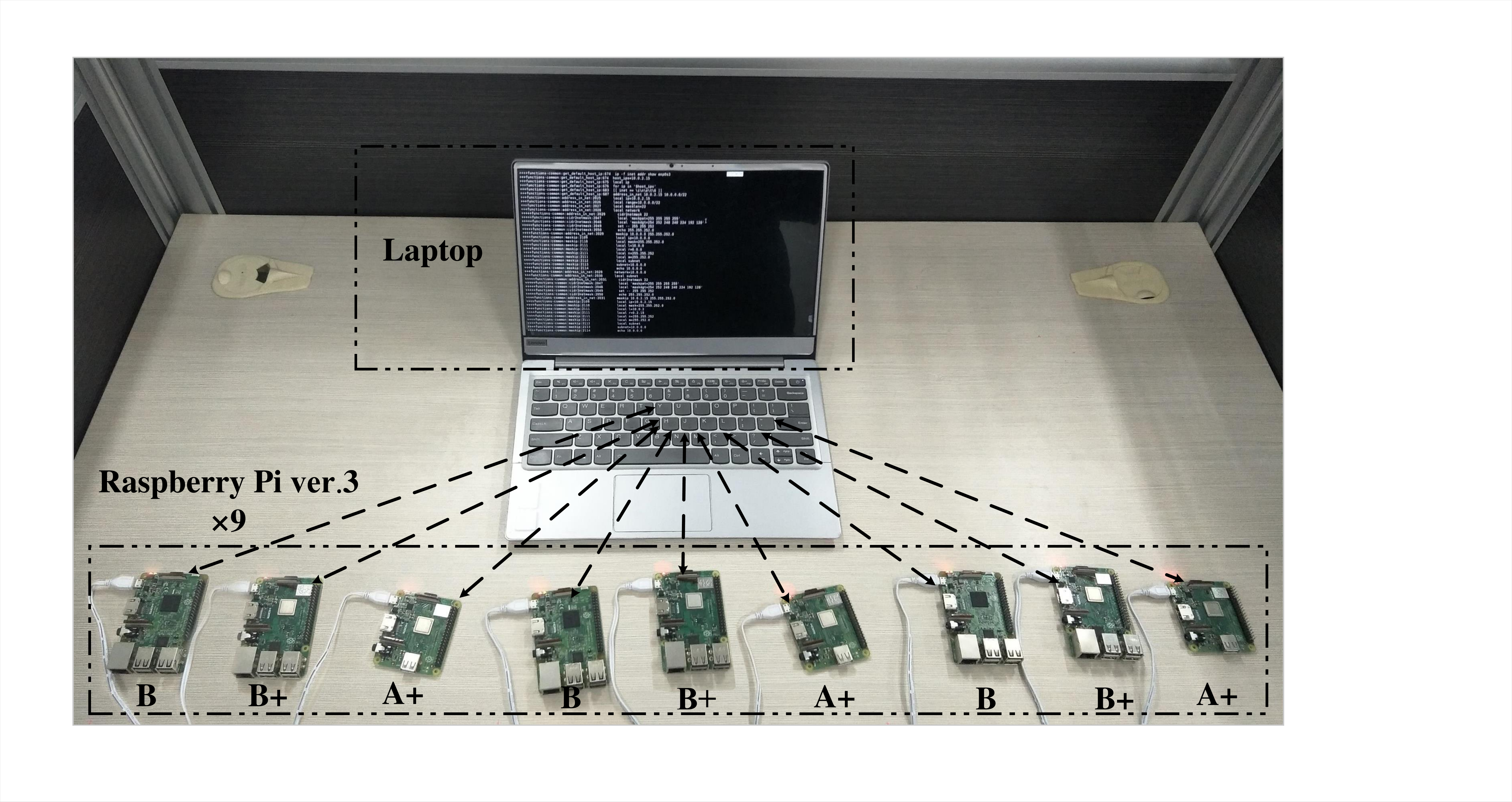}
\vspace{-0.1cm}
\caption{\footnotesize{The network topology and hardware choice in the experiments, where Raspberry Pi are with model types of A+, B, and B+.}}
\label{fig:hardware-new}
\vspace{-0.5cm}
\end{figure}

Our simulation consists of nine Raspberry Pi (version 3) and one laptop computer as shown in Fig. \ref{fig:hardware-new}, which are all interconnected via WiFi under a star network topology in an office building. This represents an edge computing environment where the computational capabilities of edge nodes are heterogeneous. The simulation parameters, e.g., the transmission and receiving energy consumption of the Raspberry Pi are both $1.42\times10^{-7}$ J/bit, the processing speed and energy consumption are $4.75\times10^{-7}$ s/bit and $3.25\times10^{-7}$ J/bit, which are based on the settings from \cite{chen2016joint}.

\subsection{Comparison Baselines and Metrics }

\textbf{Comparison Baselines.} We employ the following state-of-the-art task allocation methods as baselines. It is worth noting that the first two are some of the non-data-driven methods (e.g., synthetic method) that have been widely suggested, and the last two are the data-driven methods we proposed. 

\begin{itemize}
\item \textbf{Random Mapping (RM)} where each task is processed at different edge devices with equal probability \cite{chen2016joint}. In other words, tasks are randomly assigned.
\item \textbf{Distributed Machine Learning (DML)} distributes tasks to multiple nodes, e.g., allocating the training iteration either to edge devices or to the cloud \cite{teerapittayanon2017distributed}. 
\item \textbf{Clustered Reinforcement Learning (CRL)} conducts task allocation with our clustered reinforcement learning model.
\item \textbf{Data-driven Cooperative Task Allocation (DCTA)} leverages an SVM model to adjust the decision of the CRL model.
\end{itemize}

{\textbf{Evaluation Metrics.} From the perspective of the following metrics, we compare our proposed DCTA method with the others above state-of-the-art.}

{\textbf{1) Overall Merit (OM).} Given an allocation method, the ability to provide credible overall merit (e.g., energy saving) is crucial to all stakeholders. For interested readers, a more concrete definition of overall merit is available in previous Section~\ref{Sec:notation}.}


{\textbf{2) Processing Time (PT).} Our decision should be conducted before the deadline, the processing time we measure is the time the main device needs to partition the application and receive the output of the decision results. Formally, 
\begin{eqnarray*}
\emph{PT} =  t_s - t_c,
\end{eqnarray*}
where $t_s$ denotes the time instant when final decision is made; $t_c$ denotes the time when each experiment start.}

\textbf{3) Energy Consumption (EC).} Energy consumption is significantly critical for edge devices because most edge devices are energy-constrained. Formally, the energy consumption is defined as follows:
\begin{eqnarray*}
\emph{EC} =  \sum \limits_{p \in \bm{P}} E_p + E_t,
\end{eqnarray*}
where $E_p$ and $E_t$ denote the processing and transmission energy consumption of processor $p$, respectively.

\begin{figure*}[htpb]
    \vspace{0.2cm}
	\begin{minipage}[htpb]{0.32\linewidth}
		\centering
		\vspace{-0.4cm}
		\includegraphics[width=2.5in]{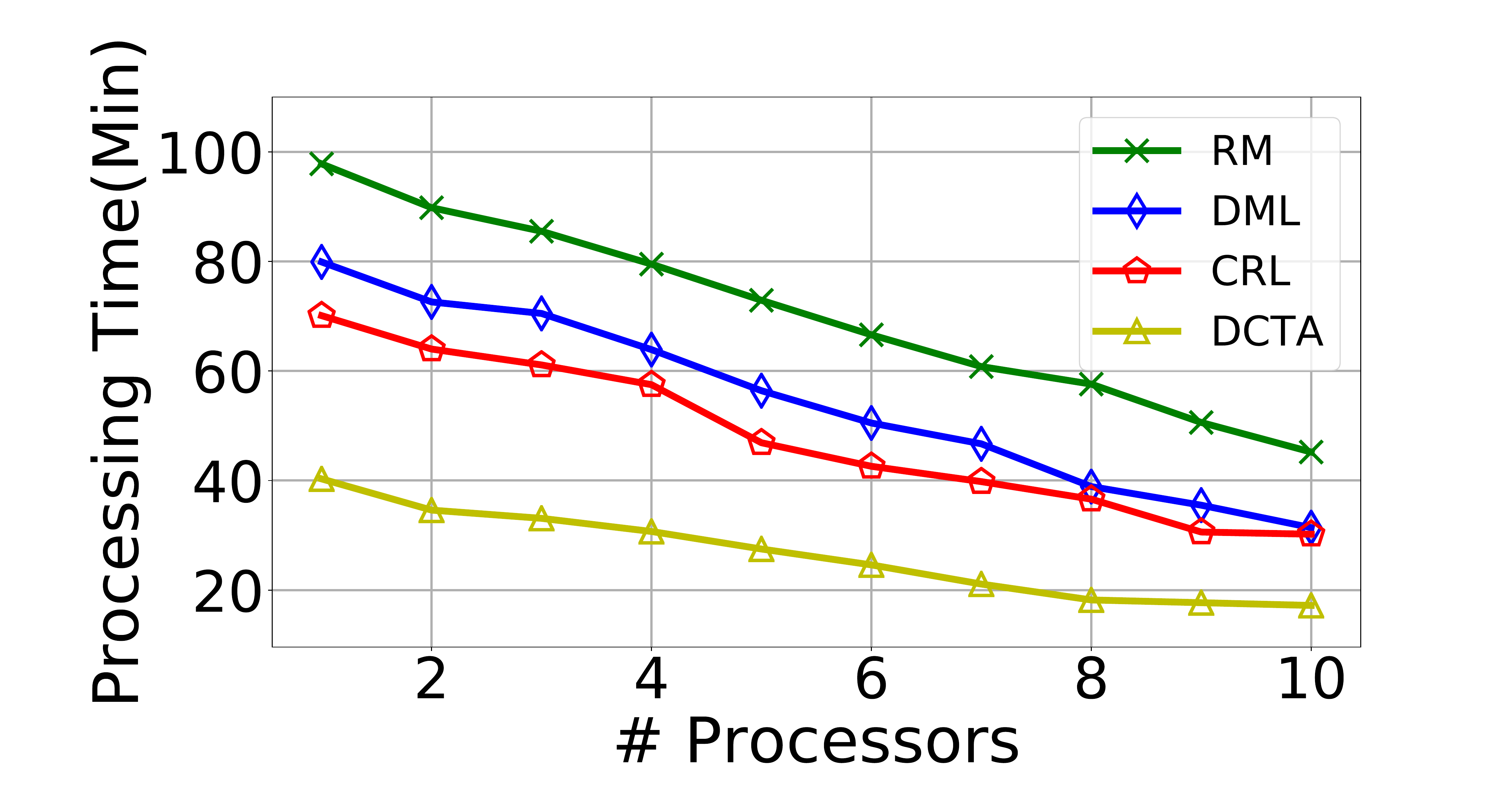}
		\vspace{-0.8cm}
		\caption{The processing time of task allocation system with different number of processors. 
        \label{fig:time-processor}}
	\end{minipage}
	\hfill
	\begin{minipage}[htpb]{0.32\linewidth}
		\centering
		\vspace{-0.5cm}
		\includegraphics[width=2.5in]{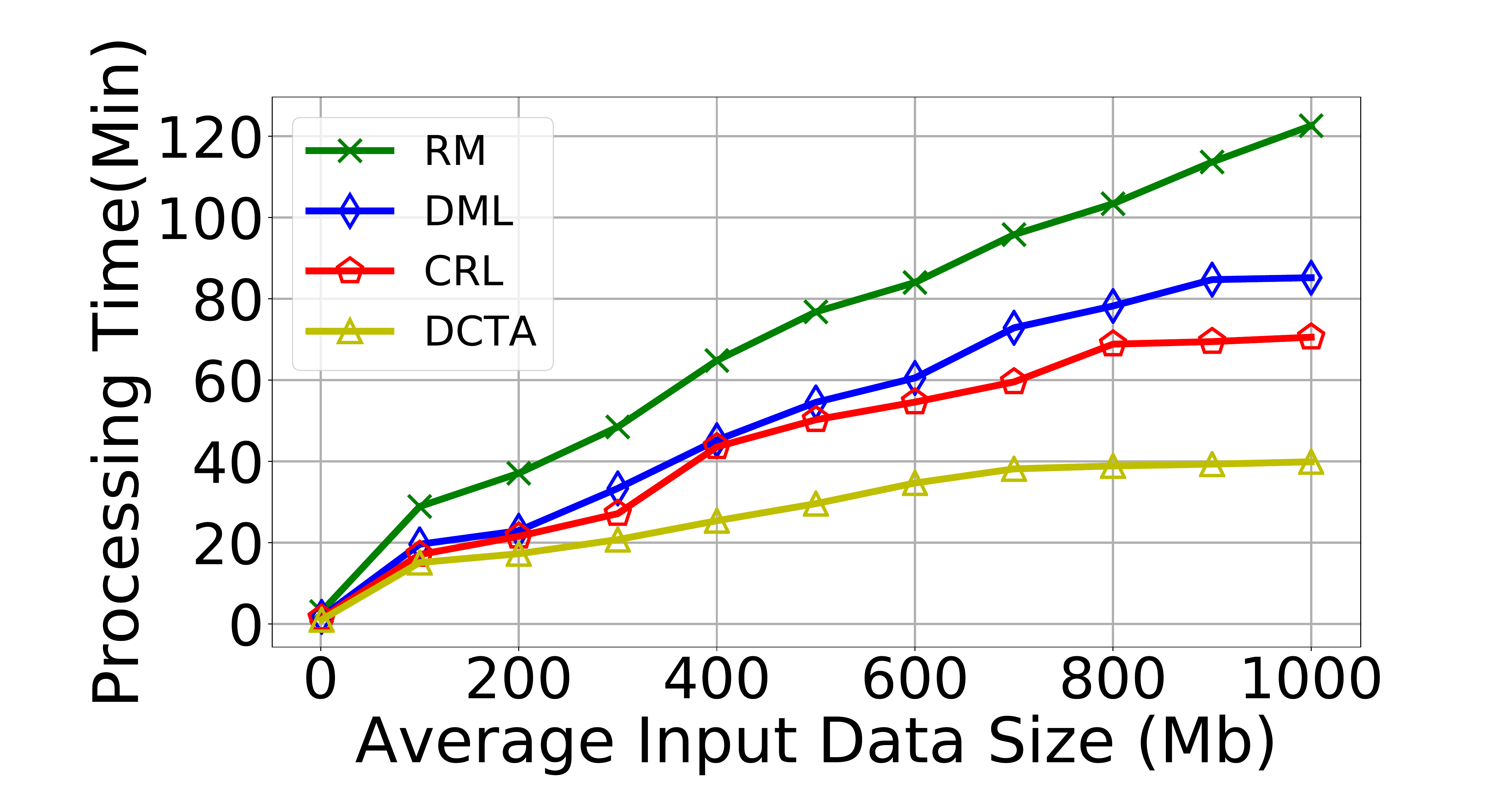}
		\vspace{-0.7cm}
		\caption{The processing time of task allocation system with different data input sizes. 
        \label{fig:time-data}}
	\end{minipage}
	\hfill
	\begin{minipage}[htpb]{0.32\linewidth}
		\centering
		\vspace{-0.5cm}
		\includegraphics[width=2.5in]{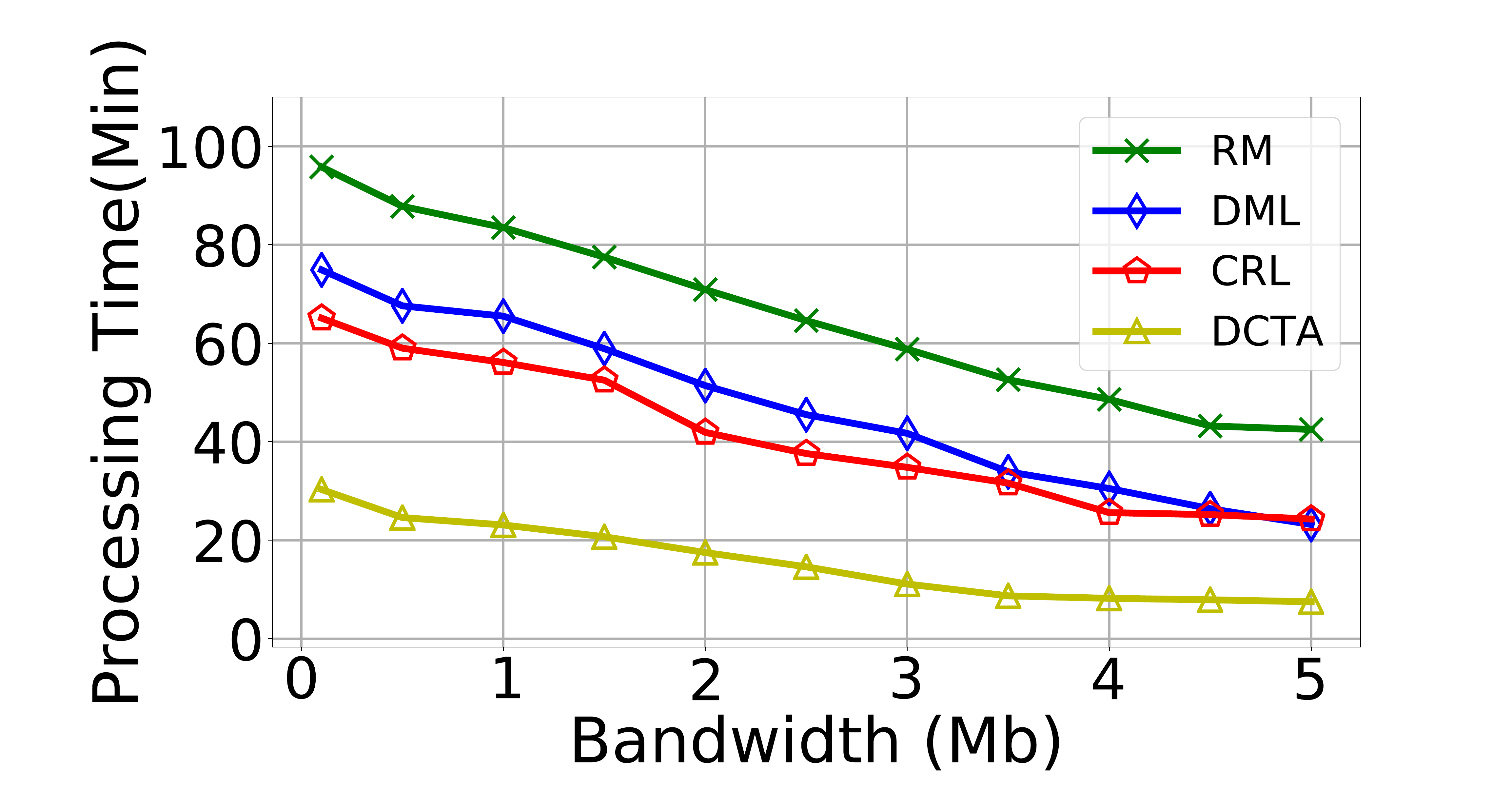}
		\vspace{-0.7cm}
		\caption{The processing time of task allocation system with different bandwidth limits. 
        \label{fig:time-bandwith}}
	\end{minipage}
	\hfill
\vspace{-0.4cm}
\end{figure*}

\subsection{Experiment Results}

\textbf{Result on Processing Time.} 
Fig. \ref{fig:time-processor} shows the processing time as a function of processors. Consistent with our intuition, as the number of processors increases, the processing time of the above methods gradually decreases. We see that DCTA can outperform RM, DML, and CRL by as much as 3.24, 2.32 and 2.01 times, respectively. On average, DCTA outperforms RM, DML, and CRL by 2.70, 2.05, and 1.80 times. That is because DCTA leverages data-driven techniques to capture the dynamic task importance and reduces the number of less important prediction tasks to perform.

Then, we compare the processing time of DCTA with that of RM, DML, and CRL for different average input data sizes. As we can see in Fig. \ref{fig:time-data}, the processing time of our DCTA is always outperformed other state-of-the-art methods. For example, our DCTA has an improvement that is 2.71, 1.83, and 1.68 times to that of RM, DML, and CRL at the average input data size of 500 Mb. That is because our DCTA obtains the importance of each task which is time-dynamic changing, and then allocates to the most suitable edge devices to execute.

Finally, Fig. \ref{fig:time-bandwith} shows the processing time as a function of network bandwidth. It is well known that network bandwidth affects the time of data transmission, and transmission time is also the main component of processing time. Thus, as the network bandwidth increases, the processing time also gradually decreases. But it is worth noting that our DCTA always outperforms RM, DML and CRL by 2.68, 1.94, and 1.71 times on average, respectively. That is mainly because our DCTA leverages data-driven techniques to capture the importance of each task and merely perform the most important tasks.

\section{Case Study: Chiller AIOps on the Edge}\label{Sec:CaseStudy}

{In this section, we focus on applying our DCTA approach to the real-world edge-computing system. We first introduce the background of one core industry AIOps system, i.e, chiller AIOps system. We then present the overview of DCTA in chiller AIOps system and briefly introduce the system architecture and main components design within our chiller AIOps system. Finally, through extensive real-world experiments, we demonstrate the superiority of our chiller AIOps system integrating the DCTA mechanism.}

\subsection{Background of Industry AIOps System}

{An important application of MTL on the edge is \emph{AIOps}. The term \emph{AIOps}~\cite{lerner18} is coined as a system that utilizes big data, machine learning and other advanced analytics to enhance IT operations, such as monitoring, automation, and service desk, with proactive, customized and dynamic insight. Data-driven analytics have been widely suggested for IT Operations Management. According to Gartner Inc., by 2022, 40\% of all large enterprises will adopt AIOps systems \cite{cappelli18}.}

{The industry AIOps system usually consists of two stages, i.e., Data-driven Multi-task Transfer Learning and Final Optimization, and they work as follows. First, when an industrial demand arrives, AIOps systems need to choose a series of data-driven prediction tasks to conduct, e.g., by using Data-driven Multi-task Transfer Learning. Second, it comes to Final Optimization. In this stage, the AIOps systems receive all the results of previous prediction tasks and conduct decisions until the decision performance, i.e., overall merit, is no longer improved.} 

{\textbf{Chiller AIOps System.} As a case study, we focus on one of the core industry AIOps system, namely, \emph{chiller AIOps system}, i.e., AIOps system conducting chiller sequencing, is deployed for one week on May, 2019, in a high-rise office building which serves more than three thousand people. A chiller is a machine that generates cooling power in commercial buildings and chiller sequencing is a significantly important operation, which aims to select run-time configurations of chillers at real-time so that the chiller AIOps system serves the time-varying cooling demand. For example, conducting chiller sequencing in a building with two chillers [0.5, 0.7] implies that chiller 1 and chiller 2 are operating at 50\% and 70\% of their maximum rated capacity, respectively. Thus, the chiller sequencing operation is to allocate the cooling load at any given time to the chillers in the most energy-efficient manner so that the overall cooling demand of the building is satisfied while at the same time the electricity consumed by the chillers is kept at a minimum~\cite{liuenergy17}. Chiller AIOps system has been studied recently to significantly improve energy efficiency in commercial buildings and this case study is conducted based on a real-world chiller operation dataset~\cite{zheng2018data}.}


\begin{figure}[t]
\centering
\vspace{-0.3cm}
\includegraphics[angle=0, width=0.3\textwidth]{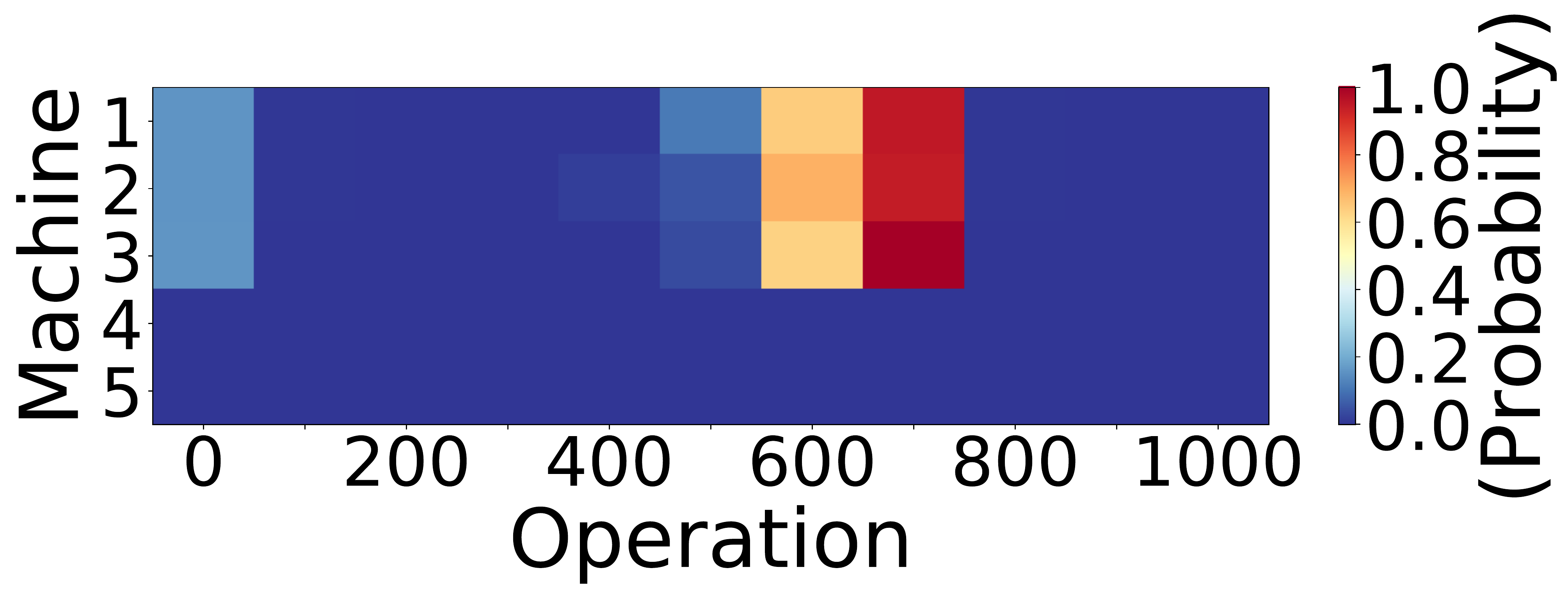}
\vspace{-0.32cm}
\caption{Probability of becoming best operation for different machines.}
\label{fig:operation}
\vspace{-0.45cm}
\end{figure}

\subsection{Overview of DCTA in Chiller AIOps System} \label{SubSec:5.2}

{As mentioned before, the efficacy of chiller sequencing control in chiller AIOps system relies heavily on the {\it run-time} performance profile of the chillers, namely the COP under different cooling load regimes. COP is a measure of the energy-efficiency of a chiller and captures the cooling power that it can output for a certain input power consumption \cite{wiki19}. Formally,
\begin{eqnarray*}
\emph{COP}_i = Q_{i} / E_{i},
\end{eqnarray*}
where $E_i$ is the electrical power consumed by chiller $i$ to deliver the required amount of cooling load $Q_i$.}

{The overall cooling load of the chiller AIOps system serves at a given time is the sum of the cooling load $Q_i$ over all chillers $i$, i.e., $Q = \sum_{i}{Q_i}$, where $Q_i=c_i \times m_i \times \Delta_T{_i}$. Here, $c_i$ is the thermal capacity of water (kJ/kg$^\circ$C), $m_i$ is the chilled water mass flow rate (kg/s) and $\Delta_T{_i}$ is the temperature difference between the returned and supplied chilled water ($^\circ$C)~\cite{liao15}. All these quantities are logged by our chiller AIOps system.}

{Reliable chiller sequencing depends on the COP across all the loading conditions for chiller $i$. However, besides the well-known fact that COP degrades over time~\cite{yu08,firdaus16}, COP also fluctuates markedly over different cooling loads and environmental conditions~\cite{zheng2018data}, which makes it exceedingly difficult to capture within an analytical model. To this end, data-driven techniques can thus play a crucial role in accurate COP prediction for improved chiller sequencing in chiller AIOps system. Specifically, a \emph{learning task} is defined as the coefficient of performance (COP) prediction of a chiller for one particular operation and works have been proposed for chiller AIOps~\cite{michopoulos07,powell13,hartman14,zheng2018data}. After COPs of operations is predicted, chiller sequencing conducted by selecting operation with the highest COP value to meet the cooling demand with the lowest electricity consumption.}

\begin{figure}[t]
\centering
\includegraphics[angle=0, width=0.48\textwidth]{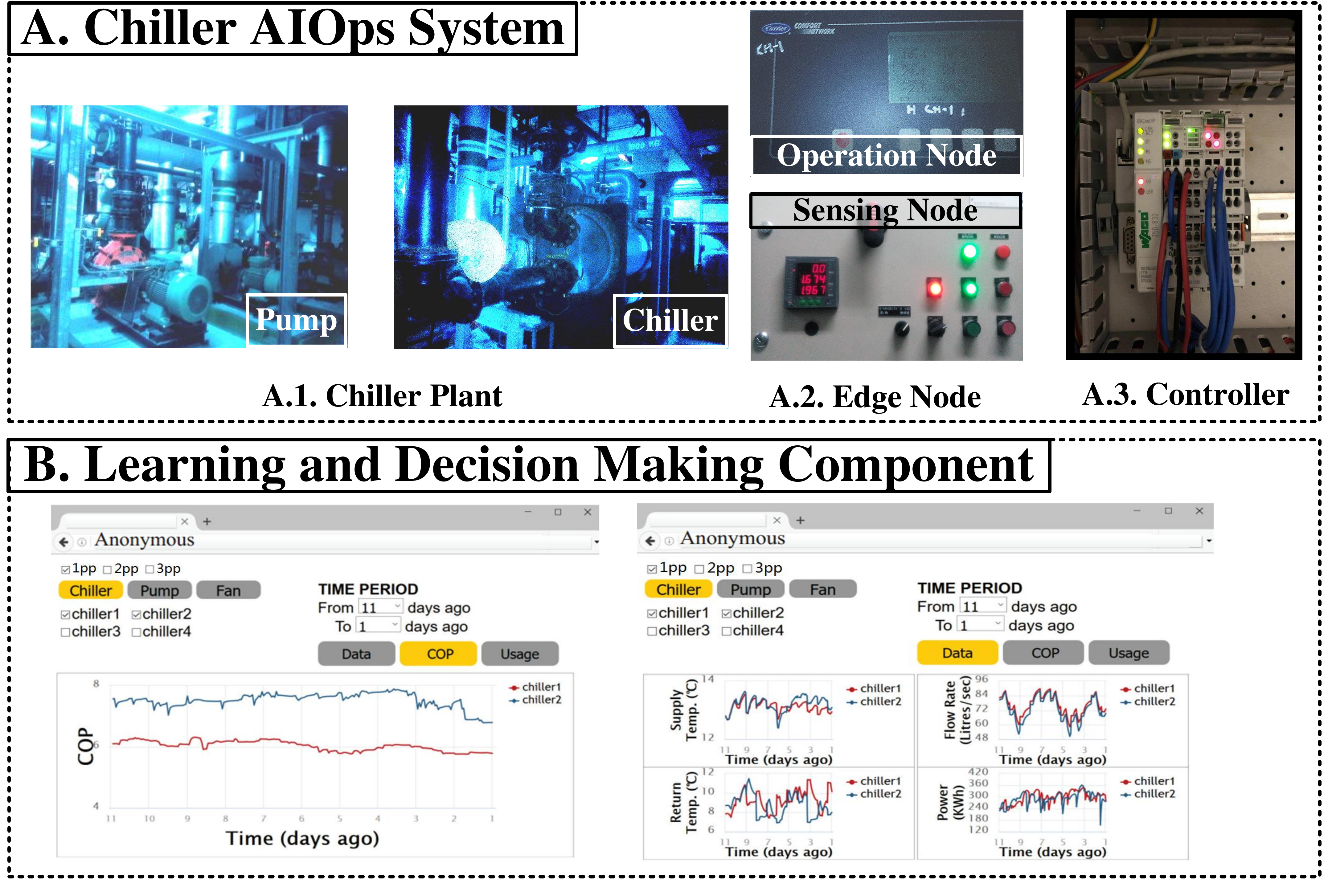}
\vspace{-0.2cm}
\caption{Chiller AIOps system on the edge.}
\label{fig:hvacReal} 
\vspace{-0.5cm}
\end{figure}

\begin{figure*}[t]
	\begin{minipage}[htpb]{0.48\linewidth}
		\centering
		\includegraphics[width=2.75in]{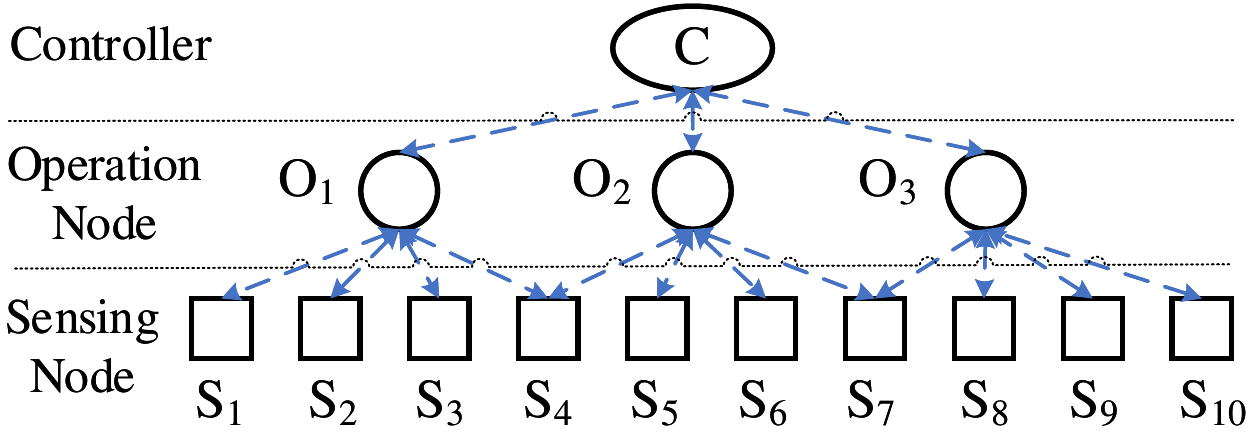}
		\caption{The network topology of chiller AIOps system on the edge. \label{fig:topology}}
	\end{minipage}
	\hfill
	\begin{minipage}[htpb]{0.48\linewidth}
		\centering
		\includegraphics[width=3.5in]{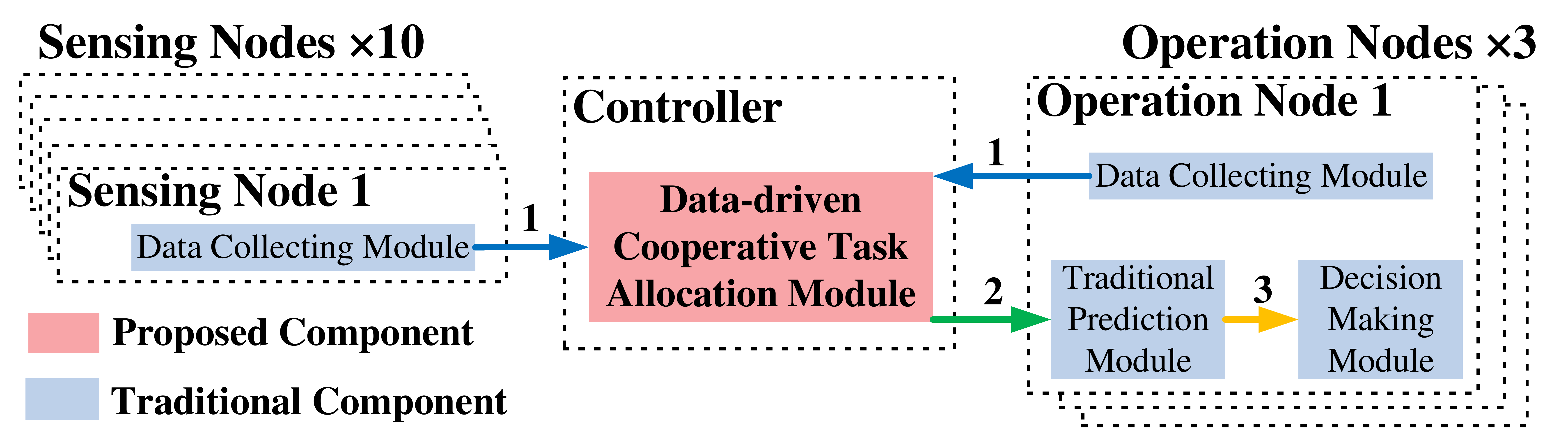}
		\vspace{-0.5cm}
		\caption{Architecture overview of chiller AIOps system on the edge. \label{fig:framework}}
	\end{minipage}
	\hfill
\vspace{-0.3cm}
\end{figure*}


{\textbf{Motivation of DCTA in Chiller AIOps.} The chiller sequencing process requires performance predicted across all possible operations. There are too many controllable parameters in the industry and the number of parameter combinations is usually huge for all possible operations. However, the chiller sequencing process is typically accompanied by time limits, e.g., two hours for chiller sequencing \cite{sun13}. A previous study indicates that blindly conducting all learning tasks leads to considerable time consumption which easily exceeds the time limits in chiller sequencing~\cite{zheng2018data}. When merely partial operations are conducted in random order and these operations fail to meet the cooling demand, the backup chiller plant would be launched and additionally consumes a large amount of electricity~\cite{yu10}. Therefore, we can conduct the proposed task allocation which assigns more important tasks to more powerful edge devices for priority execution under time limits.}

{Based on the above real-world chiller operation dataset, while in principle all COP operations (i.e., learning tasks) may be selected to conduct the chiller sequencing, in practice only a small subset of them are frequently selected in the optimal sequencing operation. The historical best operations can be computed with the sequencing optimization based on the ground truth of COP of 1460 days from 2012 to 2015. Then we can count the number of cases for each operation to be selected as the best operation and thus obtain the probability to become optimal. For example, if an operation is selected in 120 days as the best operation over the total 1460 days, its probability to become optimal is computed as 120 / 1460 = 8.22\%. Fig.~\ref{fig:operation} shows that the probability of becoming the best operation for different machines vary greatly among the exponential optional operation space. It can be seen that there merely a small portion of operations are frequently selected. Results also confirm our previous Observation 1 in Section~\ref{Sec:observation}.}


{\textbf{Task Importance in Chiller AIOps.} The key of the DCTA lies in the computation of task importance. Next, in the context of this AIOps case study, we formally present a specified formulation of the \emph{task importance} computation.} 

{As discussed in previous Definition 1 in Section~\ref{Sec:notation}, given model parameters $\bm\theta$, the task importance $\mathcal{I}_j$ can be updated using the merit function $\mathcal{H}(\cdot)$. For interested readers, a more concrete definition of $\mathcal{H}(\cdot)$ is also available in Section~\ref{Sec:notation}, where involved two following important concepts, i.e., the ideal electricity consumption $D$ and decision-making function $\mathcal{D}(\cdot)$. Specifically, as for $D$, we first find the best operation of each chiller (i.e., with the highest COP value) in each day through the historical ground truth of COP data, and then compute the electricity consumption of conducting these operations as the ideal performance.}

{Next, the decision-making function $\mathcal{D}(\cdot)$ is intrinsically solving the chiller sequencing optimization problem finding the best chiller operations combination which minimize the total electricity consumption on one day, where all time instances in one day are denoted by $T$ and each operation is conducted at time $t \in T$. Let $L_i$ denote the maximum cooling capacity of chiller $i < n$ and $S_{i, t}$ denote the partial load ratio of chiller $i$ at time $t$. Formally,
\begin{equation*}\label{Eq:decision}
    \begin{aligned}
    \mathcal{D}(\bm{\theta}) = \min  \limits_{\bm\theta}
    \sum_{t=1}^{T} \sum^n_{i=1} L_i \cdot S_{i, t} / COP_{i, t}\\ 
    \mbox{ s.t.}
    \sum^n_{i=1} Q_i >  Q_{D}\ \ and\ \ T_N \leq T_D,
    \end{aligned}
\end{equation*}
where $COP_{i, t}$ denotes the data-driven prediction performance of chiller $i$ at time $t$; $Q_i$ and $Q_{D}$ respectively denote the cooling load produced by chiller $i$ and the total cooling demand; $T_N$ and $T_D$ denote the total processing time and the deadline, respectively. More specifically, the deadline $T_D$ here means the total time length of one chiller sequencing operation, including the computation time and the mechanical switching time, computed considering both the periodic interval $t_P$ and mechanical switching time $t_M$, e.g., $T_D = \min(t_P, t_M)$ \cite{zheng2018data}.}





\subsection{Device Overview of Chiller AIOps System} \label{sec:casestudy}
{According to above, we conduct the data-driven task allocation based on the chiller AIOps system in the Pacific Place, Hong Kong, where the network topology is shown in Fig.~\ref{fig:topology}. The equipment of chillers, pumps, air-handling unit, and cooling tower differ greatly in operation, maintenance, and services. The data of each equipment in the chiller plant (Fig.~\ref{fig:hvacReal}~A.1) are captured and transmitted by 13 edge nodes, including 3 operation nodes (from the vendor of Trane, Fig.~\ref{fig:hvacReal}~A.2) conducting and recording operations, and 10 sensing nodes (from the vendor of Schneider Electric, Fig.~\ref{fig:hvacReal}~A.2) collecting sensing data. To process data from different types of equipment, we choose a centralized approach, where edge node transmits data to the controller (from the vendor of Wago, Fig.~\ref{fig:hvacReal}~A.3), and controllers are responsible for task allocation and decision making for the edge nodes. Finally, 3 operation nodes conduct data-driven COP prediction and send control sequences to devices (Fig.~\ref{fig:hvacReal}~B). Other sensing nodes without computation power are merely used to collect data.}

{Though hardware can be fully redeployed after introducing data-driven techniques~\cite{agyapong14}, for the scalability purpose, we choose an incremental deployment for the chiller AIOps system, with minimal revision for the current HVAC system. That is to say, we leverage only the current commercial off-the-shelf components and avoid deploying any additional equipment within the HVAC system. However, we may sacrifice the probability to obtain more sensing data and have even better prediction performance, if we avoid deploying additional equipment inside the local system in each building for the scalability purpose.}

\begin{figure*}[ht]
    \vspace{-0.4cm}
	\begin{minipage}[htpb]{0.32\linewidth}
		\centering
		\vspace{-0.18cm}
		\includegraphics[width=2.45in]{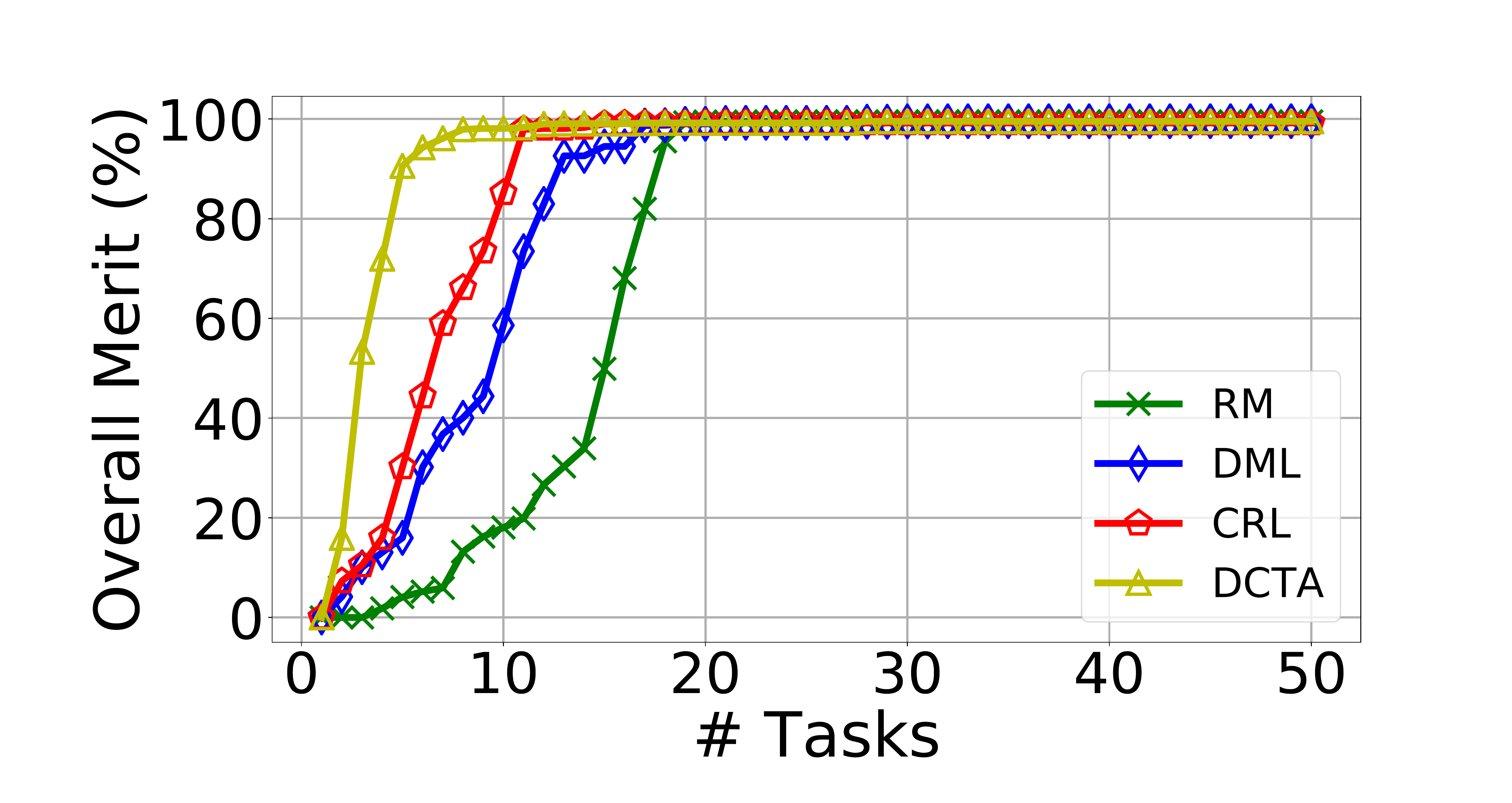}
		\vspace{-0.8cm}
		\caption{The overall merit as a function of the number of performed tasks. \label{fig:merit}}
	\end{minipage}
	\hfill
	\begin{minipage}[htpb]{0.32\linewidth}
		\centering
		\vspace{-0.2cm}
		\includegraphics[width=2.4in]{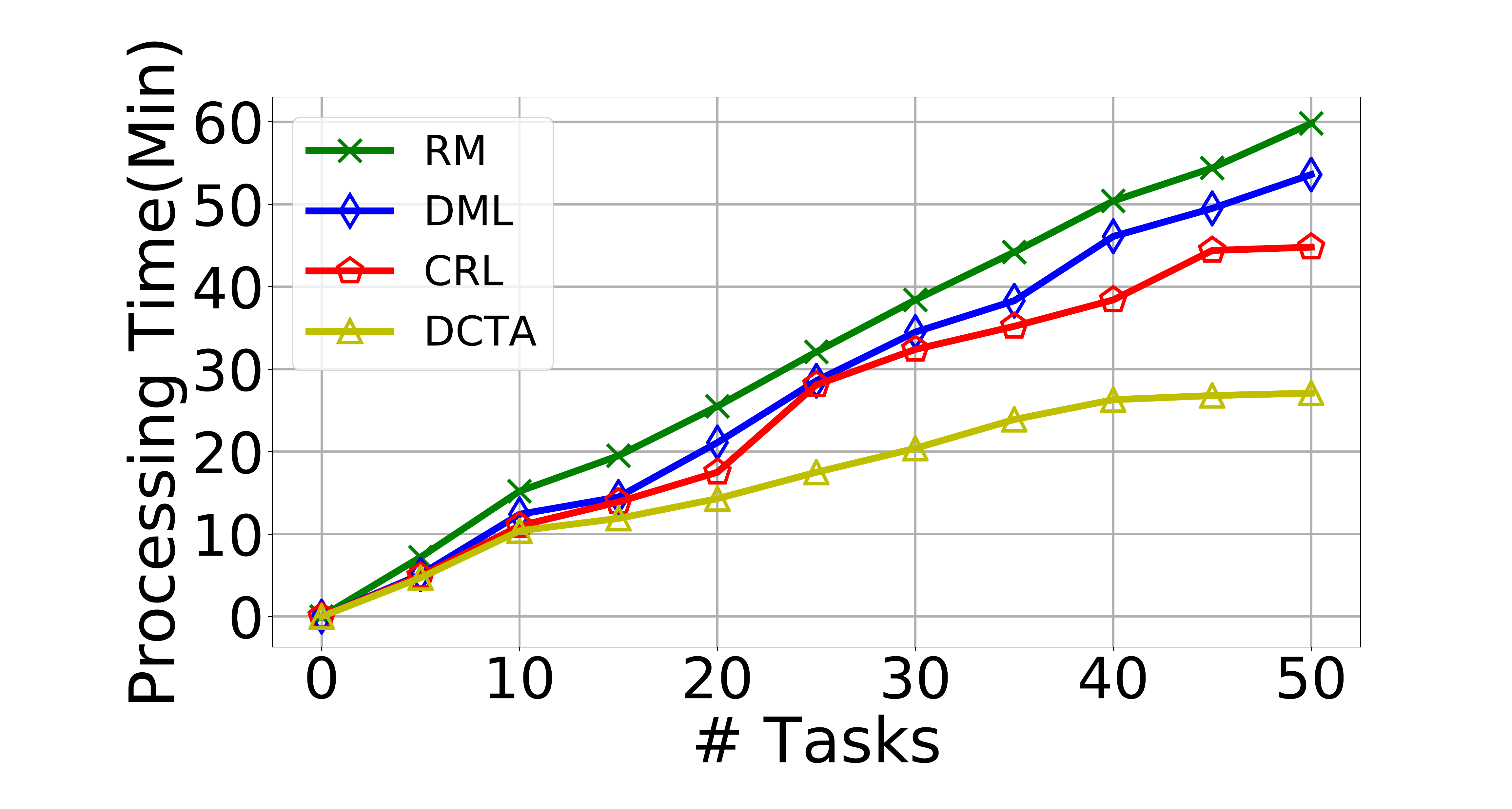}
		\vspace{-0.74cm}
		\caption{The processing time of task allocation system with different number of tasks. \label{fig:time}}
	\end{minipage}
	\hfill
	\begin{minipage}[htpb]{0.32\linewidth}
		\centering
		\vspace{0.2cm}
		\includegraphics[width=2.4in]{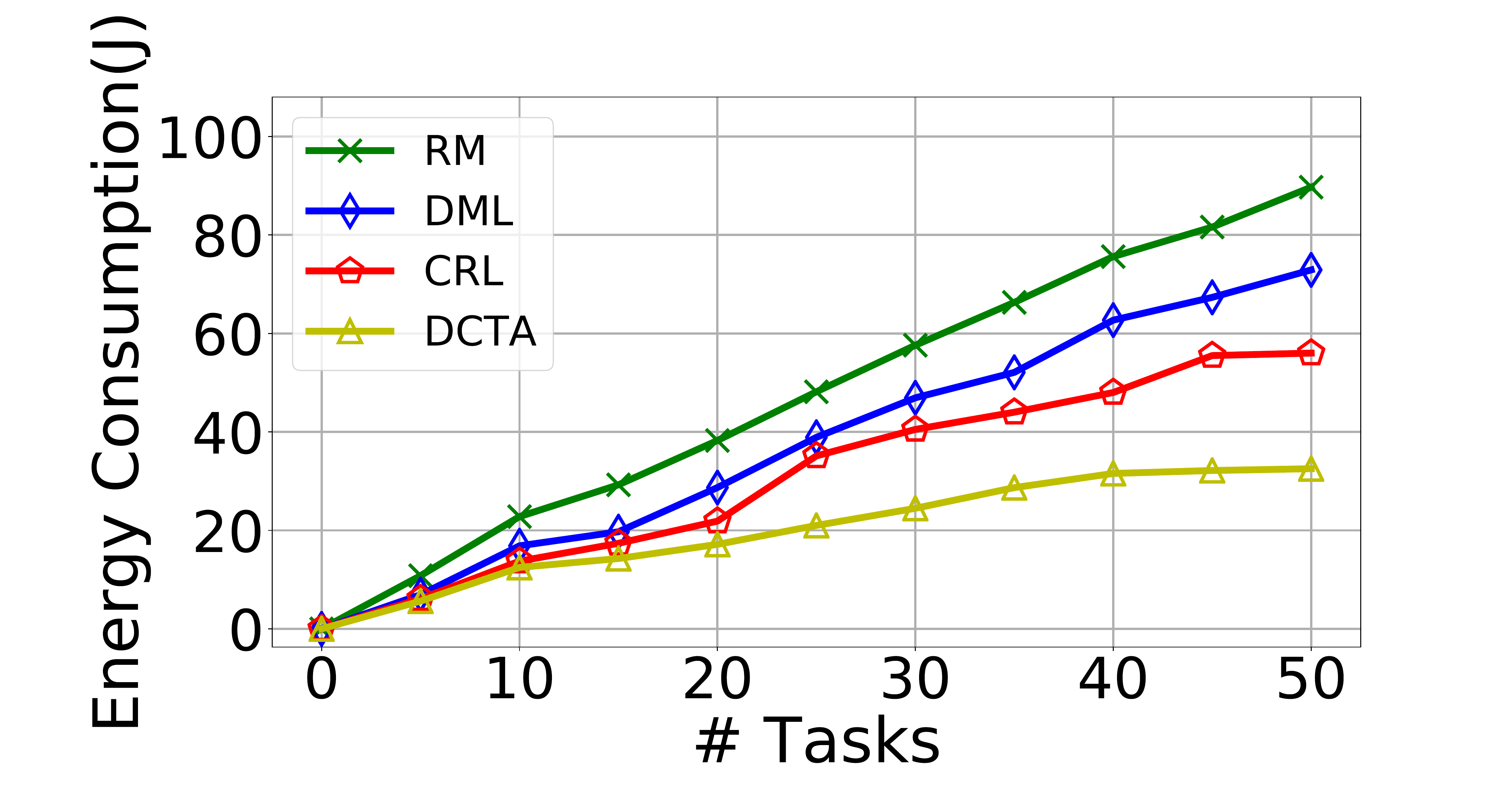}
		\vspace{-0.8cm}
		\caption{The energy consumption of task allocation system with different number of tasks. \label{fig:energy}}
	\end{minipage}
	\hfill
\vspace{-0.7cm}
\end{figure*}

\subsection{Components Design within Chiller AIOps System}

{To apply our DCTA approach to the chiller AIOps system, we also introduce the architecture overview of our chiller AIOps system, as shown in Fig. \ref{fig:framework}. The architecture contains four main modules: (1) {\em Data Collecting Module} collects the data from the surroundings for analysis. Not only the current data but also the historical data are needed to be collected. (2) {\em Data-driven Cooperative Task Allocation (DCTA) Module} captures the time-dynamic task importance and allocates tasks with data-driven techniques, which has been introduced in detail in Section \ref{Sec:Model}. (3) {\em Traditional Prediction Module} executes the data-driven prediction tasks at the edge nodes and outputs the prediction results. (4) {\em Decision Making Module} receives the prediction results from the multiple edge nodes and conducts the optimal decision which is to maximize the overall system merit.}

{The DCTA module for task allocation lies in the Controller and the design is elaborately introduced in previous Section \ref{Sec:Model}. In the following, we are to briefly introduce the design of other components, i.e., {\em data collecting module}, {\em traditional prediction module}, and {\em decision making module}, within our chiller AIOps system architecture.}

{\textbf{Data Collecting Module.} The module lies in the Sensing Nodes, e,.g, the temperature sensor or humidity sensor, which collects the data from the surroundings for analysis. There exist a common data storage problem due to the storage limitations on these edge nodes. To tackle this problem, we keep uploading data to a more powerful edge node, e.g., gateway or server, and overwrite historical data on these edge nodes when the storage is insufficient.}

{\textbf{Traditional Prediction Module.} The module lies in the Operation Nodes, e.g., gateway or router, which executes data-driven COP prediction tasks and outputs the prediction results. To ensure the accuracy of each data-driven task in the case of data scarcity on these edge nodes, we apply clustered multi-task learning approach \cite{jacob09}. It learns with training data not only from the target task, but also from other tasks, e.g., cases with similar temporal, meteorological and mechanical conditions.}

{\textbf{Decision Making Module.} The module lies in the Controller or Operation Nodes, e.g., server or gateway, which should work in an iterative optimization way. Under the circumstance, the frequency of the decision update is then critical to edge nodes network resource utilization and energy consumption. To tackle this problem, we propose an efficient algorithm to determine the frequency of decision update by analyzing the historical decision data. Specifically, we update the decision each time when an industrial demand coming. In order to reduce damage to the system, we ensure that the time interval between the two decisions can meet the needs of the system to transition from one steady state to another. As for the effects of varying this frequency, it would be an interesting future work for us to investigate the optimal frequency of decision update in industrial scenarios.}

\subsection{Experiment Results}

\textbf{Result on Overall Merit.}
{With the chiller AIOps system, we first compare the overall merit of our DCTA with that of the other state-of-the-art task allocation methods. Fig.~\ref{fig:merit} shows that, on one hand, our DCTA approach and other state-of-the-art methods can eventually achieve the same performance; on the other hand, with the same performance, our DCTA approach can greatly reduce the number of tasks performed which means significant savings in time and resources. That is because our DCTA approach is developed combined with the runtime data in the real environment and a huge amount of simulation data. In addition, it leverages the ensemble technique to avoid overfitting in non-linear modeling, which can successfully capture the system local and dynamic performance.}

\textbf{Result on Processing Time.}
{To show the potential of saving time, We compare the processing time of the state-of-the-art task allocation methods. In Fig. \ref{fig:time}, we can see that our DCTA outperforms RM, DML, and CRL by 50.2\%, 38.6\% and 30.2\%, respectively. That is because DCTA uses data-driven allocation to select the most important tasks for prediction, unlike other non-data-driven methods.}

\textbf{Result on Energy Consumption.}
{Fig. \ref{fig:energy} compares our DCTA approach with RM, DML and CRL method over a different number of tasks, in terms of Average Energy Consumption on edge devices. On average, our DCTA outperforms RM, DML and CRL by 48.4\%, 39.6\% and 31.3\%, respectively. That is because not all predictions on all operations are necessary. Our DCTA captures the top important operations and still maintains the superiority of data-driven techniques.}

\section{Related Work} \label{Sec:Related-work}

\textbf{Task Allocation} has been intensively researched in cloud computing systems ~\cite{biswas2017multi,hong2007adaptive,jiang2015reliable}. Recent years have witnessed great prospects exhibited down to the edge, e.g., from OpenCL (2008) \cite{OpenCL} to AWS IoT Greengrass (2017) \cite{AWS-Greengrass} and Microsoft Azure IoT Edge (2018) \cite{Azure-IoT-Edge}. Under edge computing, existing works on task allocation either 1) partition the machine-learning model and its input, or 2) are conducted according to different objectives.

First, task allocation in many distributed machine learning systems~\cite{hsieh17,xing15,li14,teerapittayanon2017distributed} have successfully demonstrated their effectiveness to enable big-data applications deployed on a large number of machines. 
For example, when allocating task for deep neural network (DNN), Neurosurgeon \cite{kang2017neurosurgeon} identifies a strategy in a fine-grained layer level between edge and cloud. A similar approach presented in \cite{ko2018edge} proposes a design guideline for DNN partitioning based on the layer-wise trade-off study. These methods provide the capability to accelerate the execution of a single data-driven task on the edge.

Second, existing works also consider different objectives for task allocation \cite{gaobin2019,shutong2019}. Examples include reducing the energy consumption of edge device while predefined delay constraint is satisfied \cite{cao2015energy}, finding a proper trade-off between the energy consumption and the execution delay \cite{mao2016power}, and minimizing the overall application execution cost \cite{sundar2018offloading}. A majority of these works are not designed for machine learning tasks. Nevertheless, though these techniques may consider a multi-task setting, they regard all submitted tasks as equally important, which leads to inefficient resource allocation at a task level when directly applied for MTL.  
 
Different from these works, our study investigates task allocation for multiple machine-learning tasks without knowing task priority. We capture and leverage task importance to accelerate the overall learning process, which sheds some new light on task allocation for MTL on the edge.

\textbf{Machine learning for Complicated Optimization Problems} has been successfully employed especially with time-varying parameters and complicated solutions which are repeatedly conducted \cite{samreen2016daleel,wang2018machine}. Examples include intelligent logistics \cite{li2018development}, code optimization \cite{cummins2017end,ogilvie2017minimizing,taylor2017adaptive}, task scheduling~\cite{wen2014smart,ren2017optimise,chen2018optimizing}. Our cooperative approach is closely related to ensemble learning where multiple models are used to solve an optimization problem. Ensemble learning is shown to be useful when scheduling parallel tasks \cite{emani2015celebrating} and optimizing application memory usage \cite{marco2017improving}. This work is the first attempt in applying ensemble techniques to optimize task allocation of MTL with task importance on the edge.

{\textbf{Industry AIOps.} Recent advances in machine learning have been adopted in various business applications for both individuals and enterprises, whereas the industry sector receives relatively less attention mainly due to the common issue of data scarcity, especially in the past. However, nowadays in the industry sector, the lowered cost of sensing, computing, and communications has made the impractical data-driven techniques in the late 1980s eminently practical, e.g., industrial robots, driver-less cars, and recently, energy-efficient buildings~\cite{zheng2019edge}. It is time to deliver a punch and reduce the cost using data-driven techniques on each of the industry sector. E.g., in building management systems, since the release of BLUED~\cite{anderson12} on 2012, a dataset of electricity consumption of buildings from the data analytics community of SIGKDD, various works demonstrated the need for using data analytics in building management systems. Then, in SIGKDD 2016, a data-driven study on energy breakdown in buildings reveals the huge electricity demand~\cite{batra16}. 
Nevertheless, how machine learning can be deployed is still vague in each of the industry sectors to guide mechanical operations, especially on the edge.}

\section{Discussion} \label{sec:Discussion}

Naturally, there is room for further work and possible improvements. We discuss a few points here.

\textbf{Data Scarcity on the Edge.} For industrial edge-computing applications, data scarcity often exists even though cloud storage can still cooperate for big data. The data scarcity is the result from 1) prohibitive cost or inherent difficulty in obtaining required proper training samples, 2) with respect to the application complexity and uncertainty. 
First, when considering the privacy concern, storage limitations, budget, and real-time requirements, partial or even the whole data set is not possible to be stored, transmitted and processed for the edge-computing applications, compared with that of cloud-computing applications. Meantime, due to the instability of the sensing devices, data loss also occurs frequently in some environments.
Worse still, an industrial application can be complex or highly uncertain which requires a larger amount of data. For example, many robots for text production, such as search engines or translation programs, have difficulties in finding sufficient samples for each context. The reason lies in the context of words which can result in ambiguities and there exists a huge amount of possible contexts.
Thus, we believe moves should be conducted for the data scarcity issue on the edge and we provide an edge-based MTL.

\textbf{Real-time Sensing Data.} Real-time sensing data facilitate the learning process by incorporating the run-time observations on environmental dynamics. In order to capture the run-time effect from real-time sensing data, we discuss two learning modes, i.e. the offline and online modes. First, the offline mode divides historical samples into multiple clusters in advance, e.g., using K-means. When the real-sensing data is coming, the system selects the most similar clustered samples to train and predict. Its drawback lies in the possibly low prediction accuracy due to the offline clustering. 
Second, the online mode prepares the training samples in a run-time manner by finding those which are the most similar with the real-time data, e.g., using KNN. This mode guarantees a high prediction accuracy but could lead to extra time to choose the proper training data. In this paper, we adopt the online mode to guarantee that our final decision making can be more reliable. The additional time overhead can be significantly reduced through our proposed data-driven task allocation mechanism.

{\textbf{Multi-task Assumption.} In this study, our approach is designed to tackle time-varying environments. We assume that 1) there are multiple related and indivisible machine-learning tasks, and 2) there is no strong pre- and post-dependency, which is also a prerequisite for performing multi-task transfer learning, and 3) there is not all tasks need to be learned individually from scratch to make the final decision. Thus, those cases 1) under single-task settings, or 2) under multi-task settings but with the sequential dependency between tasks, or 3) under multi-task settings but all tasks must be finished to produce the final result, are beyond the scope of this paper. It would be an interesting future work to extend our approach to those scenarios.}

\section{Conclusion} \label{Sec:Conclusion}


In this paper, we study task allocation for MTL scenarios on the edge, by introducing task importance and making the following contributions. First, we reveal that it is important to measure the impact of tasks on decision performance improvement and quantify task importance. We also observe the long-tail property of task importance, which serves as a key metric to guide task allocation, and facilitates resource saving from less important tasks. Second, we show that task allocation with task importance for MTL (TATIM) is a variant of NP-complete Knapsack problem, where the complicated computation to solve this problem needs to be conducted repeatedly under varying contexts. To solve TATIM with high computational efficiency, we propose a Data-driven Cooperative Task Allocation (DCTA) approach. Third, we conduct trace-driven simulations to evaluate the performance of the proposed DCTA approach. Extensive simulations show that our DCTA approach saves 3.24 times of processing time compared to the state-of-the-art. Finally, we add a new comprehensive real-world case study on AIOps for our DCTA approach to bridge model and practice, by proposing a new architecture and main components design within AIOps system. Extensive experiments are complemented to demonstrate the superiority, i.e., 48.4\% energy saving, of AIOps system integrating our DCTA approach. We believe that our DCTA approach offers an effective and practical mechanism for reducing the required resource associated with performing MTL on the edge.
\section*{Acknowledgements}

The authors would like to thank Zihan Lin for his valuable discussion and feedback. This work was supported in part by the NSFC under Grant 61722206 and 61761136014 (and 392046569 of NSFC-DFG) and 61520106005, in part by National Key Research \& Development (R\&D) Plan under grant 2017YFB1001703, in part by the Fundamental Research Funds for the Central Universities under Grant 2017KFKJXX009 and 3004210116, in part by the National Program for Support of Top-notch Young Professionals in National Program for Special Support of Eminent Professionals. Dan Wang's work was supported in part by RGC GRF PolyU 15210119, CRF C5026-18G, ITF UIM/363, ITF ITS/070/19FP, PolyU 1-ZVPZ, and a Huawei Collaborative Grant. All opinions, findings, conclusions and recommendations in this paper are those of the authors and do not necessarily reflect the views of the funding agencies.

\bibliographystyle{IEEEtran}
\bibliography{main.bib}

\begin{thebibliography}{10}
\providecommand{\url}[1]{#1}
\csname url@samestyle\endcsname
\providecommand{\newblock}{\relax}
\providecommand{\bibinfo}[2]{#2}
\providecommand{\BIBentrySTDinterwordspacing}{\spaceskip=0pt\relax}
\providecommand{\BIBentryALTinterwordstretchfactor}{4}
\providecommand{\BIBentryALTinterwordspacing}{\spaceskip=\fontdimen2\font plus
\BIBentryALTinterwordstretchfactor\fontdimen3\font minus
  \fontdimen4\font\relax}
\providecommand{\BIBforeignlanguage}[2]{{%
\expandafter\ifx\csname l@#1\endcsname\relax
\typeout{** WARNING: IEEEtran.bst: No hyphenation pattern has been}%
\typeout{** loaded for the language `#1'. Using the pattern for}%
\typeout{** the default language instead.}%
\else
\language=\csname l@#1\endcsname
\fi
#2}}
\providecommand{\BIBdecl}{\relax}
\BIBdecl

\bibitem{hutchinson2017overcoming}
M.~L. Hutchinson, E.~Antono, B.~M. Gibbons, S.~Paradiso, J.~Ling, and
  B.~Meredig, ``Overcoming data scarcity with transfer learning,'' \emph{arXiv
  preprint arXiv:1711.05099}, 2017.

\bibitem{yuan13}
C.~Yuan, W.~Hu, G.~Tian \emph{et~al.}, ``Multi-task sparse learning with beta
  process prior for action recognition,'' in \emph{IEEE CVPR}, 2013.

\bibitem{wu15}
Z.~Wu, C.~Valentini-Botinhao, O.~Watts \emph{et~al.}, ``Deep neural networks
  employing multi-task learning and stacked bottleneck features for speech
  synthesis,'' in \emph{IEEE ICASSP}, 2015, pp. 4460--4464.

\bibitem{emrani17}
S.~Emrani, A.~McGuirk \emph{et~al.}, ``Prognosis and diagnosis of parkinson's
  disease using multi-task learning,'' in \emph{ACM SIGKDD}, 2017.

\bibitem{zhou11}
J.~Zhou, L.~Yuan, J.~Liu, and J.~Ye, ``A multi-task learning formulation for
  predicting disease progression,'' in \emph{ACM SIGKDD}, 2011.

\bibitem{ide2017multi}
T.~Id{\'e}, D.~T. Phan, and J.~Kalagnanam, ``Multi-task multi-modal models for
  collective anomaly detection,'' in \emph{IEEE ICDM}, 2017, pp. 177--186.

\bibitem{biswas2017multi}
T.~Biswas, P.~Kuila, and A.~K. Ray, ``Multi-level queue for task scheduling in
  heterogeneous distributed computing system,'' in \emph{IEEE ICACCS}, 2017,
  pp. 1--6.

\bibitem{hong2007adaptive}
B.~Hong and V.~Prasanna, ``Adaptive allocation of independent tasks to maximize
  throughput,'' \emph{IEEE Transactions on Parallel and Distributed Systems},
  vol.~18, no.~10, pp. 1420--1435, 2007.

\bibitem{jiang2015reliable}
Y.~Jiang, Y.~Zhou, and Y.~Li, ``Reliable task allocation with load balancing in
  multiplex networks,'' \emph{ACM Transactions on Autonomous and Adaptive
  Systems (TAAS)}, vol.~10, no.~1, p.~3, 2015.

\bibitem{sundar2018offloading}
S.~Sundar \emph{et~al.}, ``Offloading dependent tasks with communication delay
  and deadline constraint,'' in \emph{IEEE INFOCOM}, 2018.

\bibitem{cao2015energy}
S.~Cao, X.~Tao \emph{et~al.}, ``An energy-optimal offloading algorithm of
  mobile computing based on hetnets,'' in \emph{IEEE ICCVE}, 2015.

\bibitem{mao2016power}
Y.~Mao, J.~Zhang, S.~Song \emph{et~al.}, ``Power-delay tradeoff in multi-user
  mobile-edge computing systems,'' in \emph{IEEE GLOBECOM}, 2016.

\bibitem{geng2018energy}
Y.~Geng, Y.~Yang \emph{et~al.}, ``Energy-efficient computation offloading for
  multicore-based mobile devices,'' in \emph{IEEE INFOCOM}, 2018.

\bibitem{chen2019}
Q.~Chen, Z.~Zheng, C.~Hu, D.~Wang, and F.~Liu, ``Data-driven task allocation
  for multi-task transfer learning on the edge,'' in \emph{IEEE ICDCS}, 2019.

\bibitem{zheng2018data}
Z.~Zheng, Q.~Chen, C.~Fan, N.~Guan, A.~Vishwanath, D.~Wang, and F.~Liu, ``Data
  driven chiller sequencing for reducing hvac electricity consumption in
  commercial buildings,'' in \emph{ACM e-Energy}, 2018.

\bibitem{yang16}
P.~Yang and J.~He, ``Heterogeneous representation learning with structured
  sparsity regularization,'' in \emph{IEEE ICDM}, 2016.

\bibitem{lin16}
K.~Lin, J.~Xu, I.~M. Baytas, S.~Ji, and J.~Zhou, ``Multi-task feature
  interaction learning,'' in \emph{ACM SIGKDD}, 2016.

\bibitem{gong12}
P.~Gong, J.~Ye, and C.~Zhang, ``Robust multi-task feature learning,'' in
  \emph{ACM SIGKDD}, 2012, pp. 895--903.

\bibitem{zhang17}
Y.~Zhang and Q.~Yang, ``Learning sparse task relations in multi-task
  learning.'' in \emph{AAAI}, 2017, pp. 2914--2920.

\bibitem{lin16Interactive}
K.~Lin and J.~Zhou, ``Interactive multi-task relationship learning,'' in
  \emph{IEEE ICDM}, 2016, pp. 241--250.

\bibitem{oyen12}
D.~Oyen, T.~Lane \emph{et~al.}, ``Leveraging domain knowledge in multitask
  bayesian network structure learning.'' in \emph{AAAI}, 2012.

\bibitem{isele16}
D.~Isele, M.~Rostami, and E.~Eaton, ``Using task features for zero-shot
  knowledge transfer in lifelong learning.'' in \emph{IJCAI}, 2016.

\bibitem{taskonomy2018}
A.~R. Zamir, A.~Sax, W.~Shen, L.~J. Guibas, J.~Malik, and S.~Savarese,
  ``Taskonomy: Disentangling task transfer learning,'' in \emph{IEEE CVPR},
  2018, pp. 3712--3722.

\bibitem{hu2018synthesize}
H.~Hu, L.~Chen, B.~Gong, and F.~Sha, ``Synthesize policies for transfer and
  adaptation across tasks and environments,'' in \emph{Advances in Neural
  Information Processing Systems}, 2018, pp. 1176--1185.

\bibitem{sax2018mid}
A.~Sax, B.~Emi, A.~R. Zamir, L.~Guibas, S.~Savarese, and J.~Malik, ``Mid-level
  visual representations improve generalization and sample efficiency for
  learning active tasks,'' \emph{arXiv preprint arXiv:1812.11971}, 2018.

\bibitem{li2018development}
X.~Li, ``Development of intelligent logistics in china,'' in \emph{Contemporary
  Logistics in China}.\hskip 1em plus 0.5em minus 0.4em\relax Springer, 2018,
  pp. 181--204.

\bibitem{iglesias2018data}
R.~Iglesias, F.~Rossi, K.~Wang, D.~Hallac, J.~Leskovec, and M.~Pavone,
  ``Data-driven model predictive control of autonomous mobility-on-demand
  systems,'' in \emph{IEEE ICRA, 2018}, pp. 1--7.

\bibitem{mnih2015human}
V.~Mnih, K.~Kavukcuoglu, D.~Silver, A.~A. Rusu, J.~Veness, M.~G. Bellemare,
  A.~Graves, M.~Riedmiller, A.~K. Fidjeland, G.~Ostrovski \emph{et~al.},
  ``Human-level control through deep reinforcement learning,'' \emph{Nature},
  vol. 518, no. 7540, p. 529, 2015.

\bibitem{bai2015information}
W.~Bai, L.~Chen, K.~Chen, D.~Han, C.~Tian, and H.~Wang, ``Information-agnostic
  flow scheduling for commodity data centers,'' in \emph{12th $\{$USENIX$\}$
  Symposium on Networked Systems Design and Implementation ($\{$NSDI$\}$ 15)},
  2015, pp. 455--468.

\bibitem{liu2017hierarchical}
N.~Liu, Z.~Li, J.~Xu, Z.~Xu, S.~Lin, Q.~Qiu \emph{et~al.}, ``A hierarchical
  framework of cloud resource allocation and power management using deep
  reinforcement learning,'' in \emph{IEEE ICDCS}, 2017.

\bibitem{chen2016joint}
M.-H. Chen, B.~Liang, and M.~Dong, ``Joint offloading decision and resource
  allocation for multi-user multi-task mobile cloud,'' in \emph{IEEE
  International Conference on Communications}, 2016, pp. 1--6.

\bibitem{teerapittayanon2017distributed}
S.~Teerapittayanon, B.~McDanel, and H.~Kung, ``Distributed deep neural networks
  over the cloud, the edge and end devices,'' in \emph{IEEE ICDCS}, 2017.

\bibitem{lerner18}
{Lerner, Andrew}. (2018) {AIOps Platforms.}
  \url{https://blogs.gartner.com/andrew-lerner/2017/08/09/aiops-platforms/}.

\bibitem{cappelli18}
{Cappelli, Will and others}. (2018) {Market Guide for AIOps Platforms.}
  \url{https://goo.gl/CHkqyN}.

\bibitem{liuenergy17}
Z.~Liu, H.~Tan, D.~Luo, G.~Yu, J.~Li, and Z.~Li, ``Optimal chiller sequencing
  control in an office building considering the variation of chiller maximum
  cooling capacity,'' \emph{Energy and Buildings}, vol. 140, pp. 430--442,
  2017.

\bibitem{wiki19}
Wikipedia. (2019) {Coefficient of performance.}
  \url{https://en.wikipedia.org/wiki/Coefficient_of_performance}.

\bibitem{liao15}
Y.~Liao, Y.~Sun, and G.~Huang, ``Robustness analysis of chiller sequencing
  control,'' \emph{Energy Conversion and Management}, vol. 103, pp. 180--190,
  2015.

\bibitem{yu08}
F.~Yu and K.~Chan, ``Optimization of water-cooled chiller system with
  load-based speed control,'' \emph{Applied Energy}, vol.~85, no.~10, pp.
  931--950, 2008.

\bibitem{firdaus16}
N.~Firdaus \emph{et~al.}, ``Chiller: Performance deterioration and
  maintenance,'' \emph{Energy Engineering}, vol. 113, no.~4, pp. 55--80, 2016.

\bibitem{michopoulos07}
A.~Michopoulos \emph{et~al.}, ``Three-years operation experience of a ground
  source heat pump system in northern greece,'' \emph{Energy and Buildings},
  vol.~39, no.~3, pp. 328--334, 2007.

\bibitem{powell13}
K.~M. Powell, W.~J. Cole \emph{et~al.}, ``Optimal chiller loading in a district
  cooling system with thermal energy storage,'' \emph{Energy}, vol.~50, pp.
  445--453, 2013.

\bibitem{hartman14}
T.~Hartman, ``All-variable speed centrifugal chiller plants,'' \emph{ASHRAE
  Journal}, vol.~56, no.~6, pp. 68--79, 2014.

\bibitem{sun13}
Y.~Sun, S.~Wang, and F.~Xiao, ``In situ performance comparison and evaluation
  of three chiller sequencing control strategies in a super high-rise
  building,'' \emph{Energy and buildings}, vol.~61, pp. 333--343, 2013.

\bibitem{yu10}
F.~Yu and K.~Chan, ``Economic benefits of optimal control for water-cooled
  chiller systems serving hotels in a subtropical climate,'' \emph{Energy and
  Buildings}, vol.~42, no.~2, pp. 203--209, 2010.

\bibitem{agyapong14}
P.~K. Agyapong \emph{et~al.}, ``Design considerations for a 5g network
  architecture,'' \emph{IEEE Communications Magazine}, vol.~52, no.~11, 2014.

\bibitem{jacob09}
L.~Jacob, J.-p. Vert, and F.~R. Bach, ``Clustered multi-task learning: A convex
  formulation,'' in \emph{NIPS}, 2009.

\bibitem{OpenCL}
{The Khronos OpenCL Working Group, "OpenCL-The open standard for parallel
  programming of heterogeneous systems"}.
  \url{https://www.khronos.org/opencl/}, January 2019.

\bibitem{AWS-Greengrass}
{AWS, "IoT Greengrass"}. \url{https://aws.amazon.com/cn/greengrass/}, 2019.

\bibitem{Azure-IoT-Edge}
{Microsoft, "Azure IoT Edge"}.
  \url{https://azure.microsoft.com/zh-cn/services/iot-edge/}, January 2019.

\bibitem{hsieh17}
K.~Hsieh, A.~Harlap, N.~Vijaykumar \emph{et~al.}, ``Gaia: Geo-distributed
  machine learning approaching lan speeds.'' in \emph{NSDI}, 2017.

\bibitem{xing15}
E.~P. Xing, Q.~Ho, W.~Dai \emph{et~al.}, ``Petuum: A new platform for
  distributed machine learning on big data,'' in \emph{ACM SIGKDD}, 2015.

\bibitem{li14}
M.~Li, D.~G. Andersen \emph{et~al.}, ``Communication efficient distributed
  machine learning with the parameter server,'' in \emph{NIPS}, 2014.

\bibitem{kang2017neurosurgeon}
Y.~Kang, J.~Hauswald, C.~Gao, A.~Rovinski \emph{et~al.}, ``Neurosurgeon:
  Collaborative intelligence between the cloud and mobile edge,'' \emph{ACM
  SIGPLAN Notices}, vol.~52, no.~4, pp. 615--629, 2017.

\bibitem{ko2018edge}
J.~H. Ko, T.~Na, M.~F. Amir, and S.~Mukhopadhyay, ``Edge-host partitioning of
  deep neural networks with feature space encoding for resource-constrained
  internet-of-things platforms,'' \emph{arXiv preprint arXiv:1802.03835}, 2018.

\bibitem{gaobin2019}
B.~Gao, Z.~Zhou, F.~Liu, and F.~Xu, ``Winning at the starting line: Joint
  network selection and service placement for mobile edge computing,'' in
  \emph{IEEE INFOCOM}, 2019.

\bibitem{shutong2019}
S.~Chen, L.~Jiao, L.~Wang, and F.~Liu, ``An online market mechanism for edge
  emergency demand response via cloudlet control,'' in \emph{IEEE INFOCOM},
  2019.

\bibitem{samreen2016daleel}
F.~Samreen, Y.~Elkhatib, M.~Rowe, and G.~S. Blair, ``Daleel: Simplifying cloud
  instance selection using machine learning,'' \emph{arXiv preprint
  arXiv:1602.02159}, 2016.

\bibitem{wang2018machine}
Z.~Wang and M.~O'Boyle, ``Machine learning in compiler optimization,''
  \emph{Proceedings of the IEEE}, no.~99, pp. 1--23, 2018.

\bibitem{cummins2017end}
C.~Cummins, P.~Petoumenos, Z.~Wang, and H.~Leather, ``End-to-end deep learning
  of optimization heuristics,'' in \emph{IEEE PACT}, 2017.

\bibitem{ogilvie2017minimizing}
W.~F. Ogilvie, P.~Petoumenos, Z.~Wang, and H.~Leather, ``Minimizing the cost of
  iterative compilation with active learning,'' in \emph{Proceedings of the
  2017 International Symposium on Code Generation and Optimization}, 2017, pp.
  245--256.

\bibitem{taylor2017adaptive}
B.~Taylor, V.~S. Marco, and Z.~Wang, ``Adaptive optimization for opencl
  programs on embedded heterogeneous systems,'' in \emph{ACM SIGPLAN Notices},
  vol.~52, no.~5, 2017, pp. 11--20.

\bibitem{wen2014smart}
Y.~Wen, Z.~Wang, and M.~F. O'boyle, ``Smart multi-task scheduling for opencl
  programs on cpu/gpu heterogeneous platforms,'' in \emph{2014 21st
  International Conference on High Performance Computing (HiPC)}, 2014, pp.
  1--10.

\bibitem{ren2017optimise}
J.~Ren, L.~Gao, H.~Wang, and Z.~Wang, ``Optimise web browsing on heterogeneous
  mobile platforms: a machine learning based approach,'' in \emph{IEEE
  INFOCOM}, 2017, pp. 1--9.

\bibitem{chen2018optimizing}
S.~Chen, J.~Fang, D.~Chen, C.~Xu, and Z.~Wang, ``Optimizing sparse
  matrix-vector multiplication on emerging many-core architectures,''
  \emph{arXiv preprint arXiv:1805.11938}, 2018.

\bibitem{emani2015celebrating}
M.~K. Emani and M.~O'Boyle, ``Celebrating diversity: a mixture of experts
  approach for runtime mapping in dynamic environments,'' in \emph{ACM SIGPLAN
  Notices}, vol.~50, no.~6, 2015, pp. 499--508.

\bibitem{marco2017improving}
V.~S. Marco, B.~Taylor, B.~Porter \emph{et~al.}, ``Improving spark application
  throughput via memory aware task co-location: A mixture of experts
  approach,'' in \emph{ACM Middleware}, 2017, pp. 95--108.

\bibitem{zheng2019edge}
Z.~Zheng, Q.~Chen, C.~Fan, N.~Guan, A.~Vishwanath, D.~Wang, and F.~Liu, ``An
  edge based data-driven chiller sequencing framework for hvac electricity
  consumption reduction in commercial buildings,'' \emph{IEEE Transactions on
  Sustainable Computing}, 2019.

\bibitem{anderson12}
K.~Anderson \emph{et~al.}, ``{BLUED:} a fully labeled public dataset for
  {Event-Based} nilm research,'' in \emph{ACM SIGKDD Workshop on SustKDD},
  2012.

\bibitem{batra16}
Batra \emph{et~al.}, ``Gemello: Creating a detailed energy breakdown from just
  the monthly electricity bill,'' in \emph{ACM SIGKDD}, 2016.

\end{thebibliography}
\vspace{-1.5cm}

\begin{IEEEbiography}[{\includegraphics[width=1in,height=1.25in,clip,keepaspectratio]{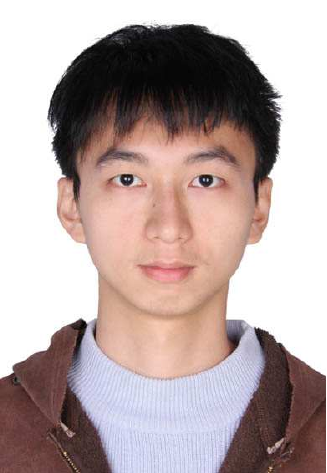}}]{Zimu Zheng} received his B.Eng. degree (Software Engineering) in South China University of Technology in 2014 and Ph.D degree (Computer Science) in the Hong Kong Polytechnic University in 2019. He is currently a research staff at Cloud BU of Huawei. Zimu has received several awards for outstanding technical contributions in Huawei. He also received the Best Paper Award of ACM e-Energy and the Best Paper Award of ACM BuildSys in 2018. His research interest lies in applied machine learning with IoT Data.
\end{IEEEbiography}

\begin{IEEEbiography}[{\includegraphics[width=1in,height=1.25in,clip,keepaspectratio]{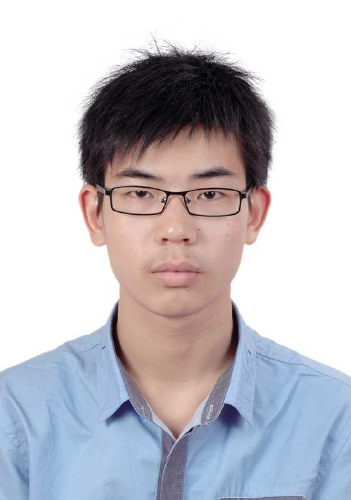}}]{Qiong Chen}
received his B.Eng. degree in School of Computer Science and Technology, Huazhong University of Science and Technology, Wuhan, China. He is currently a M.Eng. student in School of Computer Science and Technology, Huazhong University of Science and Technology. His research interests include applied machine learning and edge computing. He received the Best Paper Award of ACM International Conference on Future Energy Systems (ACM e-Energy) in 2018.
\end{IEEEbiography}

\begin{IEEEbiography}[{\includegraphics[width=1in,height=1.25in,clip,keepaspectratio]{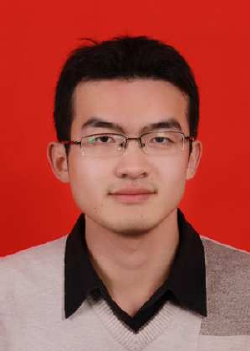}}]{Chuang Hu}
received his B.Sc degree and M.Sc degree from the Department of Computing Science, Wuhan University, China, in 2013 and 2016, respectively. He is currently a Ph.D. student in the Department of Computing, Hong Kong Polytechnic University. His research interests include networking, wireless technologies, IoT, cloud computing, and big data.
\end{IEEEbiography}

\begin{IEEEbiography}[{\includegraphics[width=1in,height=1.25in,clip,keepaspectratio]{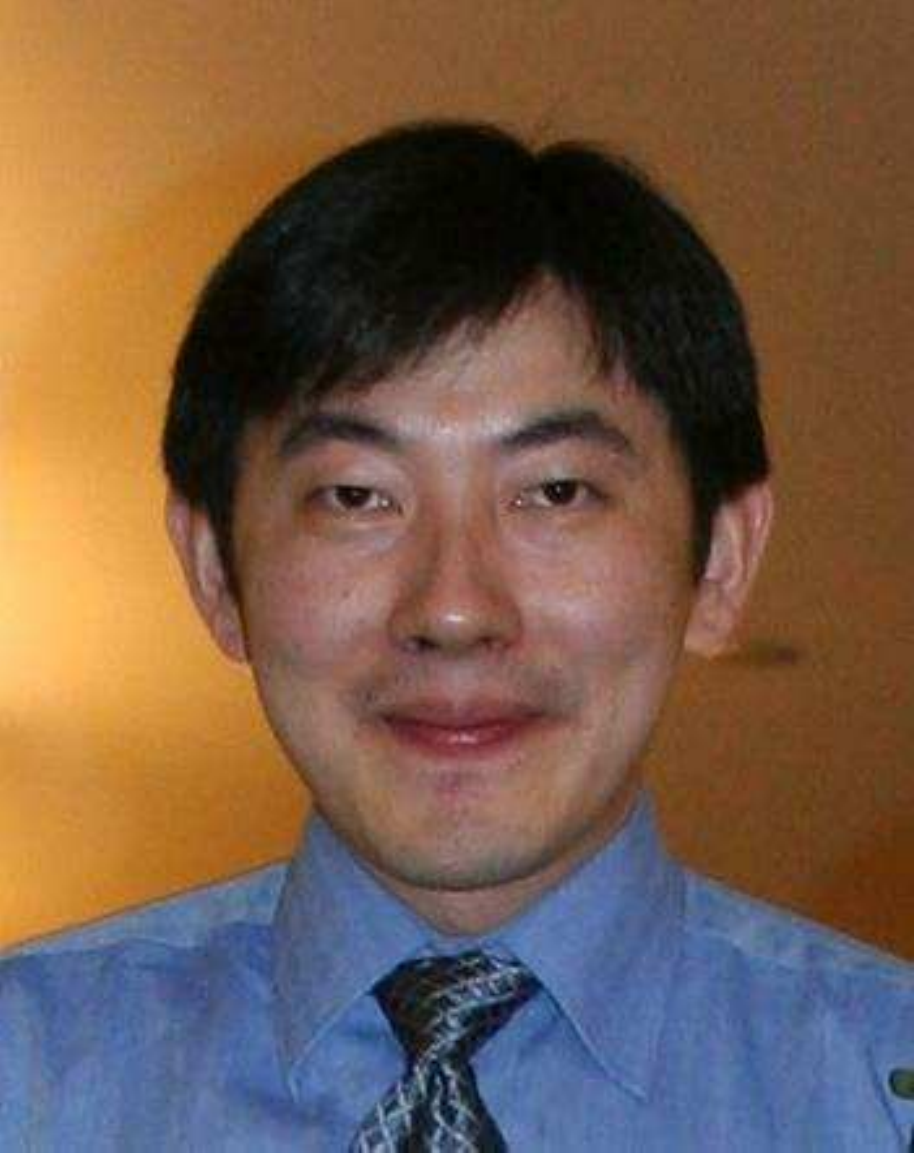}}]{Dan Wang}
(S'05, M'07, SM'13) received his B.Sc degree from Peking University, Beijing, China, in 2000, his M.Sc degree from Case Western Reserve University, Cleveland, Ohio, in 2004, and his Ph.D. degree from Simon Fraser University, Burnaby, British Columbia, Canada, in 2007, all in computer science. He is currently an associate professor at the Department of Computing, Hong Kong Polytechnic University. His research interests include network architecture and QoS, smart buildings and Industry 4.0.
\end{IEEEbiography}

\vspace{-8.8cm}
\begin{IEEEbiography}[{\includegraphics[width=1in,height=1.25in,clip,keepaspectratio]{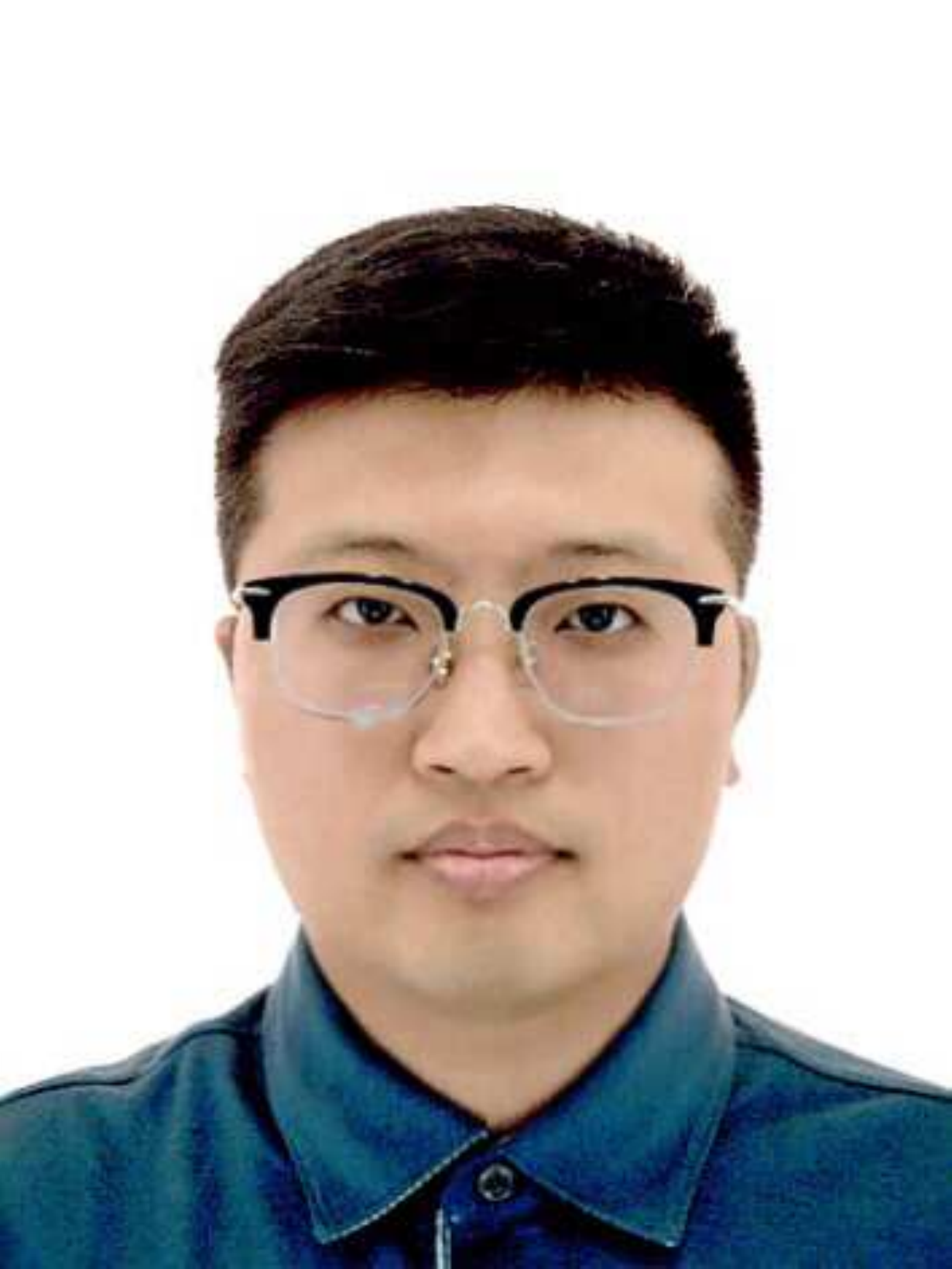}}]{Fangming Liu}
(S'08, M'11, SM'16) received the B.Eng. degree from the Tsinghua University, Beijing, and the Ph.D. degree from the Hong Kong University of Science and Technology, Hong Kong. He is currently a Full Professor with the Huazhong University of Science and Technology, Wuhan, China. His research interests include cloud computing and edge computing, datacenter and green computing, SDN/NFV/5G and applied ML/AI. He received the National Natural Science Fund (NSFC) for Excellent Young Scholars, and the National Program Special Support for Top-Notch Young Professionals. He is a recipient of the Best Paper Award of IEEE/ACM IWQoS 2019, ACM e-Energy 2018 and IEEE GLOBECOM 2011, as well as the First Class Prize of Natural Science of Ministry of Education in China.
\end{IEEEbiography}
\vspace{-1cm}

\end{document}